\documentclass[paper,10pt,nofootinbib,notitlepage,showpacs,showkeys]{revtex4}

\usepackage{graphicx}
\usepackage{amsmath}
\usepackage{amssymb}
\usepackage{dcolumn}
\usepackage{bm}
\usepackage{color}
\usepackage{dsfont}

\newcommand \beq{\begin{eqnarray}}
\newcommand \eeq{\end{eqnarray}}

\newcommand \be{\begin{equation}}
\newcommand \ee{\end{equation}}

\newcommand{\rt}[1]{{}}
\usepackage{tikz}

\begin{document}
\allowdisplaybreaks

\title{A critical look at the role of the bare parameters\\ in the renormalization of $\Phi$-derivable approximations}

\author{Urko Reinosa}
\email{reinosa@cpht.polytechnique.fr}
\affiliation{Centre de Physique Th{\'e}orique, Ecole Polytechnique, CNRS, 91128 Palaiseau Cedex, France.}

\author{Zsolt Sz{\'e}p}
\email{szepzs@achilles.elte.hu}
\affiliation{Centre de Physique Th{\'e}orique, Ecole Polytechnique, CNRS, 91128 Palaiseau Cedex, France.}

\date{\today}

\begin{abstract}
We revisit the renormalization of $\Phi$-derivable approximations from a slightly different point of view than the one which is usually followed in previous works. We pay particular attention to the question of the existence of a solution to the self-consistent equation that defines the two-point function in the Cornwall-Jackiw-Tomboulis formalism and to the fact that some of the ultraviolet divergences which appear if one formally expands the solution in powers of the bare coupling do not always appear as divergences at the level of the solution itself. We discuss these issues using a particular truncation of the $\Phi$ functional, namely the simplest truncation which brings non-trivial momentum and field dependence to the two-point function.
\end{abstract}


\maketitle

\section{Introduction}

Developed in the context of non-relativistic many-body theory \cite{LW,Baym,MdeD}, the $\Phi$-derivable approximation scheme was later generalized to relativistic quantum field theories \cite{Cornwall:vz} where it is considered a promising tool to address some of the currently open fundamental questions. This functional method relies on the two-particle irreducible (2PI) effective action, which upon a diagrammatic expansion in 2PI skeleton graphs leads to a systematically improvable approximation scheme of a given quantum field theory.  Remarkably, the global symmetries of the classical action are preserved at any order of the truncation and the solution of the self-consistent propagator equation obtained through a variational principle satisfies thermodynamical consistency \cite{Baym}. This feature makes the method particularly suitable for the calculation of thermodynamical quantities, such as the entropy \cite{Blaizot:2000fc} or the pressure \cite{Berges:2004hn}, which show a rather good convergence behavior as the order of the approximation is increased. In an out-of-equilibrium setting, the success of the 2PI evolution equations in describing the late-time dynamics of quantum fields from far-from-equilibrium initial conditions  \cite{Berges:2000ur} has opened the possibility to study a number of problems, such as preheating \cite{Berges:2002cz,Arrizabalaga:2004iw}, transport coefficients \cite{Aarts:2003bk} or topological defect formation \cite{Rajantie:2006gy,Berges:2010nk}.\\

Together with the increasing number of applications of the 2PI effective action, new insight has been gained on technical aspects regarding its renormalization \cite{vanHees:2001ik,Blaizot:2003an,Berges:2005hc}. Part of the difficulty in this context comes from the fact that the two-point function is given in terms of a self-consistent and thus non-perturbative equation, also known as the {\it gap equation}, which makes it difficult to identify and eliminate the ultraviolet (UV) divergences.  Interestingly however, part of the intuition that we have gained regarding the renormalization of $\Phi$-derivable approximations comes from perturbation theory. In fact, one can always consider a formal (partial \cite{Reinosa:2009tc} or complete \cite{Blaizot:2003an}) perturbative expansion of the solution of the gap equation which allows to unveil the structure of the corresponding UV divergences. This structure can then be expressed in terms of certain non-perturbative objects fulfilling their own self-consistent equations, in particular a four-point function obeying a Bethe-Salpeter-type equation \cite{vanHees:2001ik,Blaizot:2003an}. It follows that the renormalization procedure for $\Phi$-derivable approximations can be formulated solely in terms of these non-perturbative objects, without any reference to the formal expansion used to identify the divergences, and it is then readily applicable in practice. However, since the previous construction is based on a perturbative expansion, the only rigorous statement that can be made a priori concerning the renormalization procedure is that, if one would formally expand the solution of the gap equation in powers of the renormalized coupling, the coefficients of such an expansion would converge as the ultraviolet cut-off $\Lambda$ is taken to $\infty$. But what about the solution of the gap equation itself, before it is formally expanded? Is the renormalization procedure we referred to above sufficient to make the solution insensitive to the cut-off when the later becomes large? Numerical studies seem to indicate that this is indeed the case \cite{Berges:2004hn,Fejos:2011zq}. However, proving this fact rigorously is a difficult matter because the UV structure of the solution of the gap equation usually reveals itself after solving the equation and in general this can be achieved only numerically. Although there exist certain analytic arguments which corroborate these numerical observations \cite{vanHees:2001ik,Blaizot:2003an}, they are all based on two important assumptions. First, that the solution of the gap equation exists, if not for arbitrarily large, at least for sufficiently large $\Lambda$. Second, that the asymptotic behavior of the solution is only mildly modified with respect to that in perturbation theory, that is up to some powers of logarithms of the momentum. These two assumptions are again difficult to prove analytically and usually one needs to rely on numerical evidence. A related question is how the divergences of the formal perturbative expansion of the solution of the gap equation appear at the level of the solution itself. Since such divergences are to be absorbed in a redefinition of the bare parameters, this question can also be stated as follows: what is the actual role played by the bare parameters in the existence and the large-$\Lambda$ behavior of the solution of the gap equation?\\

In this paper we put forward a slightly different point of view towards understanding the renormalization of $\Phi$-derivable approximations, which might shed some light on the issues mentioned above. Generalizing to a momentum dependent self-energy our approach initiated in \cite{Reinosa:2011ut}, we discuss the behavior of the solution of the gap equation as the cut-off increases using a combination of numerical and semi-analytical methods, without ever relying on any perturbative expansion. This approach also addresses the question of the existence of a solution of the gap equation for arbitrary large values of $\Lambda$. Part of the originality of this work is that, in a sense, it revisits the question of renormalization of $\Phi$-derivable approximations from scratch: we shall recover known results, but from a different perspective. In particular, we point out a difference between the role of the bare mass and that of the bare coupling: if the bare mass is needed to absorb divergences of the solution of the gap equation, the role of the bare coupling is somewhat different, at least in the approximation that we consider here. We concentrate on the simplest $\Phi$-derivable approximation of the one-component $\varphi^4$ scalar field theory which introduces a non-trivial momentum dependence for the two-point function. The corresponding {\it self-energy} $\bar M^2(K)$ obeys the following gap equation:
\beq\label{eq:2loopeq}
\bar M^2(K) & = & m_0^2+\frac{\lambda_0}{2}\,\phi^2+\frac{\lambda_0}{2}\int_{|Q|<\Lambda}\bar G(Q)-\frac{\lambda_0^2}{2}\,\phi^2\mathop{\int_{|Q|<\Lambda}}_{|Q-K|<\Lambda}\bar G(Q)\bar G(Q-K)\,,
\eeq
where $m^2_0$ and $\lambda_0$ denote the bare parameters and $\bar G(Q)\equiv 1/(Q^2+\bar M^2(Q))$. In what follows, the two integrals appearing in Eq.~(\ref{eq:2loopeq}) are named respectively the {\it tadpole} integral and the {\it bubble} integral. The function $\bar M^2(K)$ is referred to as the self-energy, although it includes the tree-level bare mass. It depends on the cut-off $\Lambda$ and on the field $\phi,$ but we shall leave these dependencies implicit. The same remark applies to the propagator $\bar G(Q)$. Finally, the regularization that we have chosen\footnote{We choose here a regularization such that the norm of any momentum is limited by $\Lambda$. This regularization slightly differs from the regularization which amounts to limiting the norm of loop momenta only. In fact, only the former seems to be defined consistently as a regularization at the level of the path integral. It is also easily generalizable to smooth regulating functions, see App.~\ref{sec:reg}.} is such that the norm of any momentum, including the external momentum $K$, is less than $\Lambda$.\\

It is usually said that the self-energy $\bar M^2(K)$, solution of Eq.~(\ref{eq:2loopeq}), diverges. What is meant by this is that if one would let the cut-off $\Lambda$ grow indefinitely, keeping the bare parameters fixed, the solution $\bar M^2(K)$ of this {\it bare gap equation} would diverge. Is this really so? Note that the reason why we believe that there should be divergences comes from our experience with perturbation theory. More precisely, if we would formally expand the solution of Eq.~(\ref{eq:2loopeq}) in powers of $\lambda_0$, the coefficients of this expansion would diverge as $\Lambda\rightarrow\infty$, with divergences coming both from the tadpole and the bubble integrals. But what about the solution $\bar M^2(K)$, before it is expanded? Does it diverge? And do the possible divergences originate from both integrals? In fact, it is not even clear that the solution exists for arbitrary large values of $\Lambda$. So, how can we even start discussing divergences? We first investigate these questions in Sec.~\ref{sec:one}, where we argue that the solution of the bare gap equation exists indeed for arbitrarily large values of $\Lambda$ and diverges. Surprisingly however, it exhibits a purely quadratic divergence, that is a divergence void of multiplicative or additive logarithms, which moreover originates exclusively from the tadpole integral. Because this quadratic divergence is independent of the field, we are naturally led in Sec.~\ref{sec:two} to absorb it in a redefinition of the bare mass $m^2_0$, without redefining the bare coupling $\lambda_0$. We observe then that the solution of the corresponding {\it mass-renormalized gap equation} shows no more divergences, in contrast to what one would naively expect from the fact that the bare coupling was maintained fixed. Instead, for some values of the parameters, there exists a ``critical'' value of the cut-off above which the equation has no solution, and the very question of divergences is not well posed. We explain the origin of this critical cut-off using a ``mean-field approximation'' of the gap equation. This approximation is also used to illustrate another type of situation which could be encountered within certain $\Phi$-derivable approximations, namely the existence of a continuum limit which is however only reached for extremely large and practically inaccessible values of the cut-off. Both the existence of a critical cut-off and the very slow convergence towards a possible continuum limit, although they cannot really be referred to as divergences of the solution of the gap equation, present the same limitation than a divergent solution: they prevent the definition of a cut-off insensitive self-energy. In fact, these inconvenient features can be traced back to the presence of logarithmically divergent contributions in the mass-renormalized gap equation. These divergences can be absorbed in a redefinition of the bare coupling, but we stress the fact that these are divergences of the equation itself not of its solution and as such, the role of coupling renormalization is not to absorb divergences of $\bar M^2(K)$, at least in the approximation that we consider in this work. Coupling renormalization is implemented in Sec.~\ref{sec:three} and the corresponding {\it completely renormalized gap equation} exhibits a solution which seems to exist for arbitrarily large values of the cut-off $\Lambda$, at least in some relevant range of parameters, and converges towards a certain limit as $\Lambda\rightarrow\infty$. This limit is approached to a very good accuracy ($\propto 1/\Lambda$ when using a sharp cut-off) already for ``accessible'' values of the cut-off.

\section{Bare gap equation}\label{sec:one}
\begin{figure}[t]
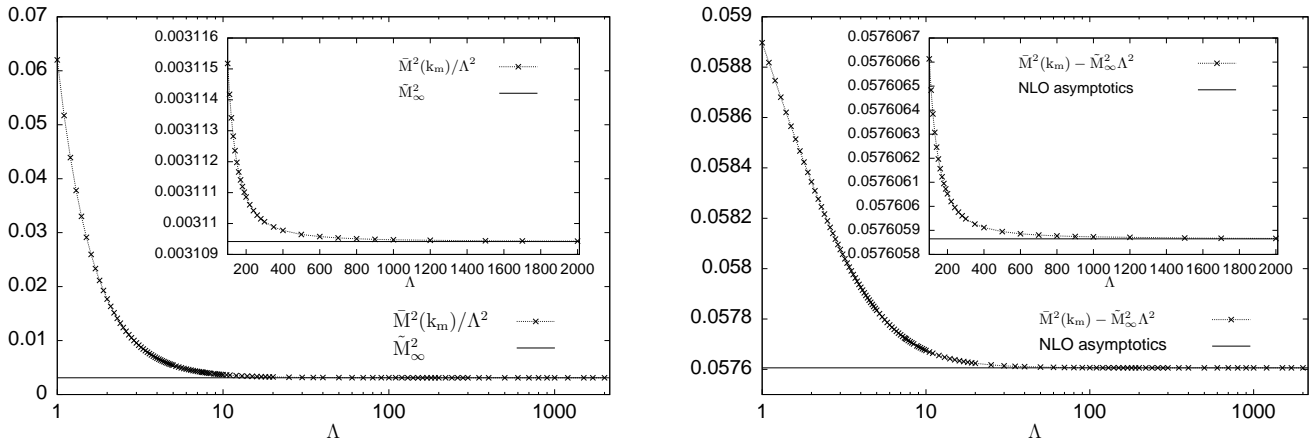

\begin{center}
\resizebox{0.45\textwidth}{!}{\input{conv1.tex}}
\hspace*{1.0cm}
\resizebox{0.45\textwidth}{!}{\input{conv2.tex}}
\caption{The left plot represents the ratio $\bar M^2(K)/\Lambda^2$ as a function of $\log_{10}\Lambda$ (and as a function of $\Lambda$ in the inset) for $|K|=k_{\rm m}=5.10^{-4}$ (the smallest momentum stored on the grid) and the comparison to the asymptotic estimate $\tilde M^2_\infty$, determined from Eq.~(\ref{eq:minf}). The right plot represents the difference $\bar M^2(k_{\rm m})-\tilde M^2_\infty\,\Lambda^2$ as a function of $\log_{10}\Lambda$ (and as a function of $\Lambda$ in the inset) and the comparison to the asymptotic estimate, as given by the right-hand-side of Eq.~(\ref{eq:subleading}). The parameters are $m^2_0=0.01$, $\lambda_0=1,$ and $\phi^2=0.1$. For a given $\Lambda$ the number of discretization points is [$100\Lambda$].
\label{fig:a}}
\end{center}
\end{figure}

Let us first assume that the bare parameters $m^2_0$ and $\lambda_0$ are kept independent of $\Lambda$. We refer to the corresponding gap equation as the {\it bare gap equation}. Our goal is to discuss the behavior of its solution $\bar M^2(K)$ as $\Lambda$ is increased for fixed $K$. To this purpose, it is convenient to consider a more general situation where some of the components of the $4$-vector $K$ grow with $\Lambda$ as well, that is $K=\Lambda\tilde K+L$, with $|\Lambda\tilde K+L|<\Lambda$. Performing the change of variables $Q=\Lambda\tilde Q$ and introducing the notations $\tilde M^2(\tilde Q)\equiv \bar M^2(Q)/\Lambda^2$, $\tilde m^2_0\equiv m_0^2/\Lambda^2$ and $\tilde\phi^2\equiv\phi^2/\Lambda^2$, the bare gap equation (\ref{eq:2loopeq}) becomes
\beq\label{eq:two}
\tilde M^2(\tilde K+L/\Lambda) & = & \tilde m_0^2+\frac{\lambda_0}{2}\,\tilde\phi^2+\frac{\lambda_0}{2}\int_{|\tilde Q|<1}\tilde G(\tilde Q)-\frac{\lambda_0^2}{2}\,\tilde\phi^2\!\!\!\mathop{\int_{|\tilde Q|<1}}_{|\tilde Q-\tilde K-L/\Lambda|<1}\tilde G(\tilde Q)\tilde G(\tilde Q-\tilde K-L/\Lambda)\,,
\eeq
with $\tilde G(\tilde Q)\equiv 1/(\tilde Q^2+\tilde M^2(\tilde Q))$. Because $\tilde m_0^2\rightarrow 0$ and $\tilde\phi^2\rightarrow 0$ as $\Lambda\rightarrow\infty$, this equation is ``compatible'' with the asymptotic behavior
\beq\label{eq:asymptotic}
\tilde M^2(\tilde K+L/\Lambda)\rightarrow \tilde M^2_\infty>0\,,
\eeq
as $\Lambda\rightarrow\infty$, for fixed $\tilde K$ and $L$, where $\tilde M^2_\infty$ fulfills the self-consistent equation\footnote{If we consider each side of this equation as a function of $\tilde M^2_\infty$, the left-hand-side increases linearly from $0$ to $\infty$, whereas the right-hand-side decreases from a strictly positive value to $0$, if $\lambda_0>0$ (which we assume to hold throughout this section and the next one). Then, because both sides of the equation depend continuously on $\tilde M^2_\infty$, there is always a unique and strictly positive solution $\tilde M^2_\infty$, when $\lambda_0>0$.}
\beq\label{eq:minf}
\tilde M^2_\infty=\frac{\lambda_0}{2}\int_{|\tilde Q|<1}\frac{1}{\tilde Q^2+\tilde M^2_\infty}\,.
\eeq
By ``compatible'' we mean that if one assumes that the asymptotic behavior (\ref{eq:asymptotic}) is obeyed by the self-energies in the right-hand-side of Eq.~(\ref{eq:two}), the later produces exactly the same asymptotic behavior for the left-hand-side of the equation. In particular, we obtain $\bar M^2(K)=\tilde M^2(K/\Lambda)\Lambda^2\sim \tilde M^2_\infty\,\Lambda^2$ that is, the solution of the bare gap equation exhibits a pure quadratic divergence at leading order in the asymptotic expansion as $\Lambda\rightarrow\infty$ for fixed $K$. Note that we have not proven that Eq.~(\ref{eq:two}) admits a solution for sufficiently large $\Lambda$. However, the fact that the contribution of the tadpole integral dominates over the contribution of the bubble integral makes this plausible. This fact is confirmed numerically for the smallest momentum stored on the grid $|K|=k_m=5.10^{-4}$ (we also take $m_0^2=0.01$ and $\phi^2=0.1$) together with the announced asymptotic behavior, see the left panel of Fig.~\ref{fig:a}. In order to save computer time, the parameters were chosen such that the asymptotic behavior (\ref{eq:asymptotic}) is observed for reasonable, that is to say, not very large values of the cut-off.\\

It is interesting to compute the first correction to the asymptotic behavior $\bar M^2(K)\sim \tilde M^2_\infty\Lambda^2$. This is done in App.~\ref{sec:subleading}. We obtain:
\beq\label{eq:subleading}
\bar M^2(K)-\tilde M^2_\infty\,\Lambda^2 & \rightarrow & \lambda_\infty\left[\frac{m^2_0}{\lambda_0}+\frac{\phi^2}{2}\right]-\frac{\lambda_0^2}{2}\,\phi^2\int_{|\tilde Q|<1}\tilde G^2_\infty(\tilde Q)+\frac{\lambda_\infty\lambda_0^2}{4}\,\phi^2\int_{|\tilde P|<1}\tilde G^2_\infty(\tilde P)\mathop{\int_{|\tilde Q|<1}}_{|\tilde Q-\tilde P|<1}\tilde G_\infty(\tilde Q)\tilde G_\infty(\tilde Q-\tilde P)\,,\nonumber\\
\eeq
with $\tilde G_\infty(\tilde Q)\equiv 1/(\tilde Q^2+\tilde M^2_\infty)$ and $1/\lambda_\infty\equiv 1/\lambda_0+(1/2)\int_{|\tilde Q|<1}\tilde G^2_\infty(\tilde Q)$. Equation (\ref{eq:subleading}) shows that there are no logarithmic divergences in $\bar M^2(K)$ as $\Lambda\rightarrow\infty$ for fixed $K$. This result is confirmed numerically, see the right panel of Fig.~\ref{fig:a}. Note also that the divergence of $\bar M^2(K)$ comes from the tadpole integral only. In contrast, if we were to consider the perturbative expansion of $\bar M^2(K)$ in powers of $\lambda_0$, the corresponding coefficients would diverge with divergences originating both from the tadpole integral and from the bubble integrals. This result illustrates that, upon resummation, some of the divergences which appear in the formal perturbative expansion of the solution of the gap equation, do not appear in the solution itself. Does this mean that there is no need to absorb these perturbative divergences? This is one of the questions we shall investigate in the next sections.

\section{Mass-renormalized gap equation}\label{sec:two}
The quadratic divergence $\tilde M^2_\infty\Lambda^2$ discussed in the previous section does not depend on the field $\phi$ and we can then try to ``absorb'' it by adjusting the bare mass $m^2_0$, which then becomes a function of $\Lambda$ and keeping the coupling $\lambda_0$ fixed.\footnote{Due to the presence of the term $\lambda_0\phi^2/2$ in Eq.~(\ref{eq:2loopeq}), renormalizing the coupling $\lambda_0$ would mean that the divergence of $\bar M^2(K)$ contains a terms proportionnal to $\phi^2$, which it has not, see Eq.~(\ref{eq:asymptotic})} As already mentioned, the divergence originates from the tadpole integral. However, because the latter depends on the field $\phi$ through the solution $\bar M^2(Q)$, it cannot be completely absorbed in a redefinition of $m^2_0$. Still, we shall see below that, in order to remove the divergence, it is enough to choose $m^2_0=-(\lambda_0/2)[\int_{|Q|<\Lambda} G_0(Q)\,+\,\mbox{finite part}]$ with $G_0(Q)\equiv 1/(Q^2+m^2)$ and $m^2$ some fixed renormalized mass. The bare mass is defined up to an additive ``finite part'' which we can choose such that the renormalization condition $\bar M^2=m^2$ at $\phi^2=0$ is fulfilled.\footnote{For $\phi^2=0$, the self-energy becomes momentum independent and we do not need to specify a particular momentum in the renormalization condition. Other choices of renormalization points are possible. In particular, if one is interested in discussing the broken phase, one should rather impose a renormalization condition at a given non-vanishing value of the field, or at vanishing field but non-zero temperature.} This amounts here to the particular choice:
\beq\label{eq:m0}
m_0^2=m^2-\frac{\lambda_0}{2}\int_{|Q|<\Lambda}G_0(Q)\,.
\eeq
Plugging this expression into the bare gap equation, we end up with
\beq\label{eq:2loop2}
\bar M^2(K) & = & m^2+\frac{\lambda_0}{2}\,\phi^2+\frac{\lambda_0}{2}\int_{|Q|<\Lambda}\Big[\bar G(Q)-G_0(Q)\Big]-\frac{\lambda_0^2}{2}\,\phi^2\mathop{\int_{|Q|<\Lambda}}_{|Q-K|<\Lambda}\bar G(Q)\bar G(Q-K)\,,
\eeq
which we refer to as the {\it mass-renormalized gap equation}. Note that, because we have modified the dependence of the equation with respect to $\Lambda$, we need to analyze once again the behavior of the solution of Eq.~(\ref{eq:2loop2}) as $\Lambda$ increases. Using the same rescaling and notations as in the previous section, we obtain
\beq\label{eq:rescaled}
\tilde M^2(\tilde K+L/\Lambda) & = & \tilde m^2+\frac{\lambda_0}{2}\,\tilde \phi^2+\frac{\lambda_0}{2}\int_{|\tilde Q|<1}\Big[\tilde G(\tilde Q)-\tilde G_0(\tilde Q)\Big]-\frac{\lambda_0^2}{2}\,\tilde \phi^2\!\!\!\!\!\mathop{\int_{|\tilde Q|<1}}_{|\tilde Q-\tilde K-L/\Lambda|<1}\!\!\!\!\!\tilde G(\tilde Q)\tilde G(\tilde Q-\tilde K-L/\Lambda)\,,
\eeq
where $\tilde m^2=m^2/\Lambda^2$ and $\tilde G_0(\tilde Q)\equiv 1/(\tilde Q^2+\tilde M^2(\tilde Q))$. This equation is compatible with $\tilde M^2(\tilde K+L/\Lambda)\equiv\bar M^2(\Lambda\tilde K+L)/\Lambda^2$ becoming smaller and smaller as $\Lambda$ increases for fixed $\tilde K$ and $L$ (this will be confirmed numerically in what follows) and supports then the fact that the choice (\ref{eq:m0}) for the bare mass eliminates the quadratic divergence. Notice however that the question of the existence of a solution for any value of $\Lambda$ is now more subtle than in the case of the bare gap equation because the contribution of the subtracted tadpole integral does not necessarily dominate over the contribution of the bubble integral. We shall now study more in detail what happens to the solution of the mass-renormalized gap equation as $\Lambda$ is increased.

\subsection{Effect of the bubble integral}
Before discussing the mass-renormalized equation (\ref{eq:2loop2}), let us consider a simpler equation where we focus on the effect of the bubble integral only, that is:
\beq\label{eq:2loop3}
\bar M^2(K) & = & m^2_\phi-\frac{\lambda_0^2\phi^2}{2}\mathop{\int_{|Q|<\Lambda}}_{|Q-K|<\Lambda}\bar G(Q)\bar G(Q-K)\,,
\eeq
where $m^2_\phi\equiv m^2+(\lambda_0/2)\,\phi^2$ and $\lambda_0^2\phi^2$ are considered as independent parameters. Such a simplification makes possibly sense when the field $\phi$ dominates over the mass $m$. Its main purpose here is to introduce, in a simpler context, the arguments and tools that will be used in order to deal with Eq.~(\ref{eq:2loop2}). Moreover, some of the properties of the solution of Eq.~(\ref{eq:2loop2}) will be visible here already.

\subsubsection{Numerical solution}
\begin{figure}[htbp]
\begin{center}
\resizebox{0.6\textwidth}{!}{\input{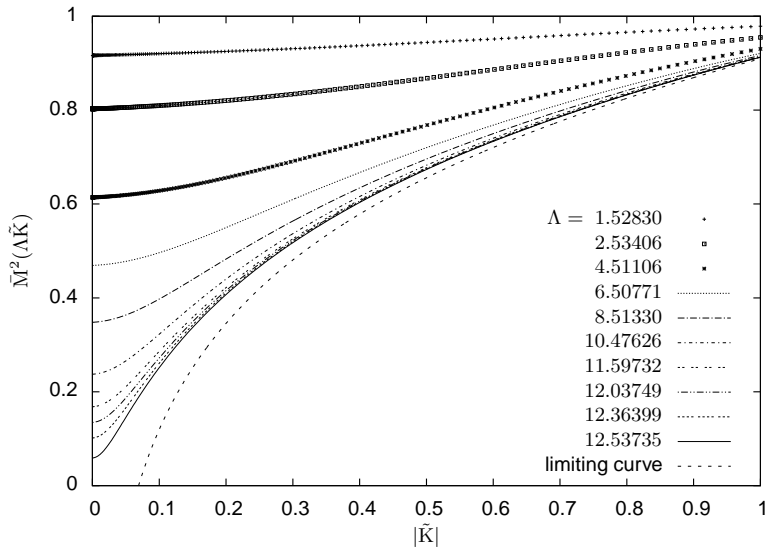}}
\caption{Evolution with $\Lambda$ of the solution $\bar M^2(K)$ of Eq.~(\ref{eq:flow}) plotted as a function of the rescaled momentum $|\tilde K|=|K|/\Lambda$. Lower curves correspond to increasing values of $\Lambda$. The flow cannot be continued above a certain critical cut-off $\Lambda_{\rm c}$. The plain curve represents the solution obtained as $\Lambda\rightarrow\Lambda_{\rm c}^-$. The dashed curve represents the limiting function (\ref{eq:approach}), which is approached by $\bar M^2(\Lambda\tilde K)$, at least for $|\tilde K|$ close to $1$ and if $\Lambda_{\rm c}$ is large enough, see the text for details. The parameters are $m^2_\phi=1$ and $\lambda^2_0\phi^2=50,$ the smallest momentum stored on the grid is $k_{\rm m}=7.5\,10^{-4}$ and the number of discretization points is $175.$
\label{fig:aa}}
\end{center}
\end{figure}

For reasons that will become clear below, we solve Eq.~(\ref{eq:2loop3}) using a differential method, that is by increasing the cut-off  $\Lambda$ continuously from the initial value $\Lambda=0$ at which the solution of Eq.~(\ref{eq:2loop3}) is known and given by $m^2_\phi$. An ``infinitesimal'' change from $\Lambda$ to $\Lambda+\delta\Lambda$ is encoded in an evolution or {\it flow} equation which is more conveniently derived for $\bar M^2(\Lambda\tilde K)$ at $\tilde K$ fixed.\footnote{If $\Lambda\neq 0$, it is completely equivalent to consider $\bar M^2(K)$ for $|K|<\Lambda$ or $\bar M^2(\Lambda\tilde K)$ for $|\tilde K|<1$.} In fact, after a simple change of variables, we can write
\beq\label{eq:rescaled2}
\bar M^2(\Lambda \tilde K)=m^2_\phi-\frac{\lambda_0^2\phi^2}{2}\,\mathop{\int_{|\tilde Q|<1}}_{|\tilde Q-\tilde K|<1}\tilde G(\tilde Q)\tilde G(\tilde Q-\tilde K)\,,
\eeq
where we note that the integration domain does not involve the cut-off $\Lambda$ anymore. Then, taking a derivative with respect to $\Lambda$ at fixed $\tilde K$, we arrive at the equation
\beq\label{eq:flow}
\partial_\Lambda \bar M^2(\Lambda\tilde K)=\lambda_0^2\phi^2\mathop{\int_{|\tilde Q|<1}}_{|\tilde Q-\tilde K|<1}\tilde G^2(\tilde Q)\tilde G(\tilde Q-\tilde K)\left[\frac{1}{\Lambda^2}\partial_\Lambda\bar M^2(\Lambda\tilde Q)-\frac{2}{\Lambda^3}\bar M^2(\Lambda\tilde Q)\right],
\eeq
which can be solved from the initial condition\footnote{This limit is obtained from Eq.~(\ref{eq:rescaled2}) for fixed $\tilde K$ using a
consistency argument.} $\lim_{\Lambda\rightarrow 0}\,\bar M^2(\Lambda\tilde K)=m^2_\phi$ using a Runge-Kutta algorithm combined with a linear solver based on LU decomposition (see App.~\ref{sec:flow} for details). Note that Eq.~(\ref{eq:flow}) is compatible with $\lim_{\Lambda\rightarrow 0}\,\partial_\Lambda\bar M^2(\Lambda\tilde K)=0$. The flow is then flat at initialization. For
practical purposes it is then more convenient to choose a non-zero, but small value for the initial $\Lambda$. We have chosen 
$\Lambda_{\rm init}=2 k_{\rm m},$ where $k_{\rm m}$ is the value of the smallest momentum stored on the grid, which is
kept fixed in a given run. The result of integrating the flow equation is shown in Fig.~\ref{fig:aa} for a particular choice of the parameters $m^2_\phi$ and $\lambda^2_0\phi^2$. Each curve corresponds to $\bar M^2(K)$ for a given value of the cut-off $\Lambda$, plotted as a function of the rescaled momentum $|\tilde K|=|K|/\Lambda$.\\

We first observe that as $\Lambda$ increases the self-energy $\bar M^2(\Lambda\tilde K)$ becomes insensitive to the cut-off in an increasing range of $|\tilde K|,$ starting from the highest value $|\tilde K|=1$. This can be understood easily from Eq.~(\ref{eq:rescaled2}), which is compatible with $\tilde M^2(\tilde Q)$ becoming smaller and smaller as $\Lambda$ increases and thus with $\bar M^2(\Lambda\tilde K)$ approaching 
\beq\label{eq:approach}
m^2_\phi-\frac{\lambda_0^2\phi^2}{2}\mathop{\int_{|\tilde Q|<1}}_{|\tilde Q-\tilde K|<1}\frac{1}{\tilde Q^2}\frac{1}{(\tilde Q-\tilde K)^2}\,.
\eeq
This is confirmed in Fig.~\ref{fig:aa}. Of course, since the integral in (\ref{eq:approach}) diverges logarithmically as $\tilde K\rightarrow 0$, we expect the agreement between $\bar M^2(\Lambda\tilde K)$ and (\ref{eq:approach}) to be the best for $|\tilde K|$ close to $1$ and to extend over a wider range of $|\tilde K|$ as $\Lambda$ is increased, which is also visible in Fig.~\ref{fig:aa}. As we shall see in a moment, there exists an upper value of the cut-off which cannot be overpassed. This explains why in Fig.~\ref{fig:aa}, where this upper value is only slightly larger than $12.5$ ($m^2_\phi=1$ and $\lambda_0^2\phi^2=50$), the agreement between $\bar M^2(\Lambda\tilde K)$ and (\ref{eq:approach}) is only effective in the region where $|\tilde K|$ is really close to $1$.\\

\begin{figure}[htbp]
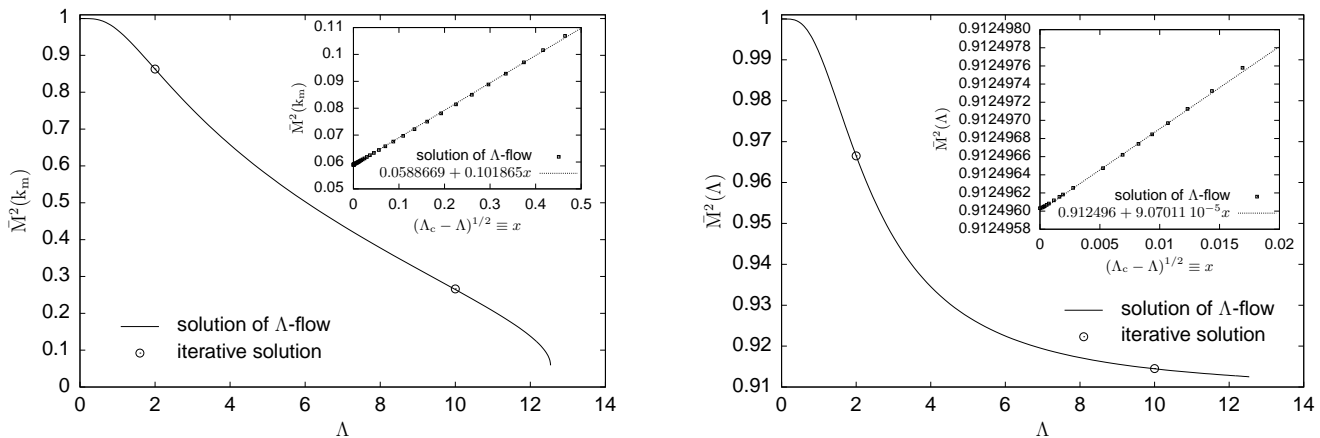

\begin{center}
\resizebox{0.45\textwidth}{!}{\input{L-flows_1st_bin_no_tad.tex}}
\hspace*{1.0cm}
\resizebox{0.45\textwidth}{!}{\input{L-flows_last_bin_no_tad.tex}}
\caption{Left plot: evolution with $\Lambda$ of the self-energy $\bar M^2(\Lambda\tilde K)$ evaluated for $|\tilde K|=0$ (numerically $|\tilde K|=k_{\rm m}/\Lambda$). Right plot: evolution with $\Lambda$ of the self-energy $\bar M^2(K)$ evaluated for the highest $|\tilde K|=1$. The parameters are $m^2_\phi=1$ and $\lambda^2_0\phi^2=50$, while $k_{\rm m}=7.5\,10^{-4},$ and the number of discretization points is $175$ for the solution of the flow and [$400\Lambda/3$] for the iterative solution.
\label{fig:bbb}}
\end{center}
\end{figure}

The previous remarks do not apply to the neighborhood of $\tilde K=0$ where $\bar M^2(\Lambda\tilde K)$ remains sensitive to the cut-off $\Lambda$. In order to study more precisely what happens in this small momentum region, in the left plot of Fig.~\ref{fig:bbb} we consider
the evolution with $\Lambda$ of the self-energy evaluated for $|\tilde K|=0$,\footnote{For numerical convenience we rather evaluate $\bar M^2(K)$ at $|K|=k_{\rm m},$ which should be regarded as our numerical approximation of $\bar M^2(0)$.} which we denote by $\bar M^2=\bar M^2(0)$ from now on. We observe that, after a relatively smooth evolution, the derivative of $\bar M^2$  with respect to $\Lambda$ becomes infinite (numerically at least) for some ``critical'' value $\Lambda_{\rm c}$ and the mass $\bar M^2$ approaches a non zero value $\bar M^2_{\rm c}\neq 0$. A fit in the vicinity of $\Lambda_{\rm c}$ reveals that the singularity is well approximated by
\beq\label{eq:behavior}
\bar M^2-\bar M^2_{\rm c}\propto\big(\Lambda_{\rm c}-\Lambda\big)^{1/2}\,.
\eeq
The flow equation (\ref{eq:flow}) suggests that this singular behavior ``propagates'' to non-zero values of $\tilde K$, as we confirm in the right plot of Fig.~\ref{fig:bbb} for the highest rescaled momentum available on our grid, that is $|\tilde K|=1$. Note also that, despite the fact that the smallest value of the self-energy, obtained at zero momentum, decreases with increasing $\Lambda$ (due to the bubble integral contribution), the flow stops before the propagator develops a pole at zero-momentum.\\

The presence of a singularity in the flow does not necessarily mean that there is no solution to Eq.~(\ref{eq:2loop3}) above $\Lambda_{\rm c}$. It only means that if a solution exists for such values of $\Lambda$, it cannot be accessed by integrating the flow equation from $\Lambda=0$, without any additional information. For a given $\Lambda$, we can also solve Eq.~(\ref{eq:2loop3}) using iterations from an initial (constant or perturbative) ansatz for the self-energy or from a solution to the gap equation obtained at a different value of the cut-off. The results of such an iterative procedure are represented in the plots of Fig.~\ref{fig:bbb} for some values of $\Lambda$ and they compare pretty well with those obtained from the flow equation. We observe that the iterative procedure fails around $\Lambda_{\rm c}$ as well. This again does not necessarily mean that there is no solution to Eq.~(\ref{eq:2loop3}) above $\Lambda_{\rm c}$ but only that if a solution exists, it cannot be reached within the iterative approach from the initial ansatz that we have  considered. In order to better grasp the origin of the singularity, we shall now consider a kind of ``mean-field approximation'' for Eq.~(\ref{eq:2loop3}) which, because it will decouple the different momenta, will turn the original integral equation (\ref{eq:2loop3}) into a certain number of simple numeric equations. The mean-field approximation will not only allow us to understand analytically the origin of  the singularity, it will also allow us to argue that there is indeed no solution to Eq.~(\ref{eq:2loop3}) above $\Lambda_{\rm c}$. A similar approximation will be used later to discuss the properties of the original mass-renormalized equation (\ref{eq:2loop2}).\\

Before proceeding further with the construction of a mean-field approximation, we add a few remarks concerning the numerical implementation. On the one hand, since the integration of the flow equation (\ref{eq:flow}) involves the (time consuming) numerical resolution of a linear system, we are much more limited concerning the number of discretization points in the case of the flow approach, than in the iterative one. Consequently, the latter provides a more accurate solution, even though discrepancies with the flow approach only start to appear very close to $\Lambda_{\rm c}$. On the other hand, the iterative method is not the best choice for approaching $\Lambda_{\rm c}$, since the slightest overpassing of $\Lambda_{\rm c}$ will only be visible after an important number of iterations have been considered and leads thus to an important slowing down in the process of finding $\Lambda_{\rm c}$. What is actually observed, by monitoring the change of $\bar M^2(0)$ between two consecutive iterative steps, is that the iterative method seemingly converges in its first stage, but after a large number of iterations the procedure clearly diverges. By increasing the precision of the numerical integration and the number of discretization points we could convince ourselves that it is unlikely that this feature is the consequence of error accumulation. In contrast, the integration of the flow equation with adaptive step-size is such that the flow remains always below $\Lambda_{\rm c}$. There is also a numerical slowing down as we approach the singularity, but it is globally less important than with the iterative approach, because none of the generated data is wasted.

\subsubsection{Mean-field approximation}

\begin{figure}[htbp]
\begin{center}
\resizebox{0.55\textwidth}{!}{\input{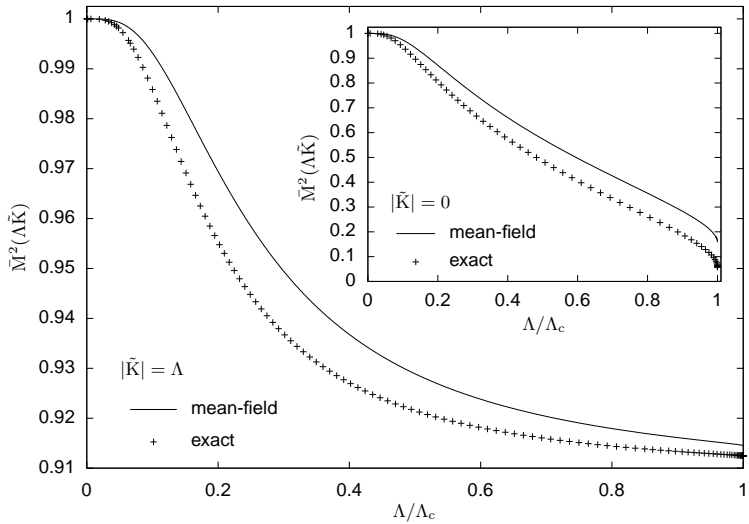}}
\caption{A comparison between the mean-field (lines) and exact (points) solution of the respective equations (\ref{eq:2loop3}) and (\ref{eq:toy})-(\ref{eq:toy3}) for $\bar M^2(\Lambda\tilde K)$ as a function of the reduced cut-off $\Lambda/\Lambda_{\rm c},$ at $|\tilde K|=\Lambda$ and $|\tilde K|=0$ (inset). For parameters 
$m^2_\phi=1$ and $\lambda^2_0\phi^2=50$ the critical cut-off is $\Lambda_{\rm c}=12.54$ in the exact case and $\Lambda_{\rm c}=9.35$ in
the mean-field case.
\label{fig:c}}
\end{center}
\end{figure}

Let us first consider the small $|\tilde K|$ region. As we already argued and as we checked numerically, $\tilde M^2(\tilde Q)$ becomes smaller and smaller as the cut-off increases. Then, the integral in Eq.~(\ref{eq:rescaled2}) in the region of small $|\tilde K|$ is dominated by small values of $|\tilde Q|$ and it makes sense to consider the following approximation for the zero momentum self-energy
\beq\label{eq:toy}
\bar M^2\equiv\bar M^2(0)=m^2_\phi-\frac{\lambda_0^2\phi^2}{2}\int_{|Q|<\Lambda}\frac{1}{(Q^2+\bar M^2)^2}\,,
\eeq
obtained from Eq.~(\ref{eq:2loop3}) by setting $K=0$ and replacing the self-energy appearing in the bubble integral by its value at $Q=0$. For non-zero values of $|K|$, we consider the following approximation
\beq\label{eq:toy3}
\bar M^2(K)=m^2_\phi-\frac{\lambda_0^2\phi^2}{2}\mathop{\int_{|Q|<\Lambda}}_{|Q-K|<\Lambda}\frac{1}{Q^2+\bar M^2}\frac{1}{(Q-K)^2+\bar M^2(K)}\,,
\eeq
where we have replaced the integral by the contributions obtained when one of the propagators reaches the smallest mass, that is $\bar M^2=\bar M^2(0)$.\footnote{The self-energy appears to be a monotonously increasing function of $|K|$, see also the discussion in App.~\ref{sec:evaluation}.} There are two such contributions, one corresponding to $Q=0$ the other to $Q+K=0$ which should be added if we were really restricting the integration domain. But because we continue integrating over the whole domain, the two contributions need to be averaged. However, due to the particular UV regularization that we have considered, the two contributions over which we average are identical and we finally end up with Eq.~(\ref{eq:toy3}). As announced, we have now a certain number of decoupled equations parametrized by $K$. All equations depend on $\bar M^2$, which needs to be determined first from Eq.~(\ref{eq:toy}). This approximation can certainly not allow us to access the actual solution of Eq.~(\ref{eq:2loop3}) and the actual value of $\Lambda_{\rm c}$. We are however only interested in the properties of $\bar M^2(K)$ as $\Lambda$ increases and those seem to be pretty well captured by the approximation, as we can see in the comparison depicted in Fig.~\ref{fig:c}. Although this is not directly visible on all curves, they all present a singularity of the type (\ref{eq:behavior}) for some critical cut-off $\Lambda_{\rm c}$. For identical values of the parameters, the value of the critical cut-off is different between the ``exact'' equation and its mean field approximation. However, note that the corresponding evolutions with $\Lambda$ are rather close to each other when plotted against the reduced cut-off $\Lambda/\Lambda_{\rm c}$.

Let us now use the mean-field approximation to understand analytically the behavior of $\bar M^2(\Lambda\tilde K)$ as $\Lambda$ increases. We shall focus on Eq.~(\ref{eq:toy}), since the appearance of a singularity in the flow of the zero-momentum self-energy triggers the appearance of a singularity in the flow of $\bar M^2(\Lambda \tilde K)$ for any other $\tilde K$. Our goal is to discuss how the solutions of Eq.~(\ref{eq:toy}) evolve with $\Lambda$. To this purpose, we rewrite this equation as $0=f_\Lambda(\bar M^2)$ with
\beq\label{eq:f}
f_\Lambda(M^2)\equiv-M^2+m^2_\phi-\frac{\lambda_0^2\phi^2}{2}\int_{|Q|<\Lambda}\frac{1}{(Q^2+M^2)^2}\,,
\eeq
and study the shape and the zeros of $f_\Lambda(M^2)$ as $\Lambda$ increases. For $\Lambda=0$, $f_0(M^2)=-M^2+m^2_\phi$ and Eq.~(\ref{eq:toy}) has one solution only: $\bar M^2=m^2_\phi$. To treat the case $\Lambda>0$, we consider the first and second derivatives of $f_\Lambda(M^2)$ with respect to $M^2$:
\beq
f'_\Lambda(M^2) & = & -1+\lambda_0^2\phi^2\int_{|Q|<\Lambda}\frac{1}{(Q^2+M^2)^3}\,,\label{eq:fp}\\
f''_\Lambda(M^2) & = & -3\lambda_0^2\phi^2\int_{|Q|<\Lambda}\frac{1}{(Q^2+M^2)^4}\label{eq:fpp}\,.
\eeq
Because $f''_\Lambda(M^2)<0$ , $f'_\Lambda(M^2)$ decreases strictly from $f'_\Lambda(0^+)=\infty$ (the integral in Eq.~(\ref{eq:fp}) is infrared divergent) to $f'_\Lambda(\infty)=-1$ (the same integral is suppressed for $M^2\gg\Lambda^2$). If we denote by $\bar M^2_{\rm e}(\Lambda)$ the only value of $M^2$ at which $f'_\Lambda(M^2)$ vanishes, we conclude that $f_\Lambda(M^2)$ has a maximum for $M^2=\bar M^2_{\rm e}(\Lambda)$. More precisely, it increases strictly from $f_\Lambda(0^+)=-\infty$ (the integral in Eq.~(\ref{eq:f}) is infrared divergent) to $f_\Lambda(\bar M^2_{\rm e}(\Lambda))$ and then decreases towards $f_\Lambda(\infty)=-\infty$ (the same integral is suppressed for $M^2\gg\Lambda^2$). It follows that the number of solutions of the equation $0=f_\Lambda(\bar M^2)$ depends on the sign of $f_\Lambda(\bar M^2_{\rm e}(\Lambda))$. To discuss this sign, note that by definition $0=f'_\Lambda(\bar M^2_{\rm e}(\Lambda))$. It follows then that
\beq
\frac{d}{d\Lambda}\bar M^2_{\rm e}(\Lambda)=-\frac{1}{f''_\Lambda(\bar M^2_{\rm e}(\Lambda))}\frac{\partial f'_\Lambda}{\partial\Lambda}=-\frac{\lambda_0^2\phi^2}{8\pi^2f''_\Lambda(\bar M^2_{\rm e}(\Lambda))}\frac{\Lambda^3}{(\Lambda^2+\bar M^2_{\rm e}(\Lambda))^3}>0\,,
\eeq
and
\beq
\frac{d}{d\Lambda}f_\Lambda(\bar M^2_{\rm e}(\Lambda))=\frac{\partial f_\Lambda}{\partial\Lambda}=-\frac{\lambda_0^2\phi^2}{16\pi^2}\frac{\Lambda^3}{(\Lambda^2+\bar M^2_{\rm e}(\Lambda))^2}<0\,.
\eeq
Thus $\bar M^2_{\rm e}(\Lambda)$ increases strictly with $\Lambda$ whereas $f_\Lambda(\bar M^2_{\rm e}(\Lambda))$ decreases strictly with $\Lambda$. Let us now determine the extremal values of $\bar M^2_{\rm e}(\Lambda)$ and $f_\Lambda(\bar M^2_{\rm e}(\Lambda))$.  After performing the integral in Eq.~(\ref{eq:fp}), the equation defining $\bar M^2_{\rm e}(\Lambda)$ reads
\beq\label{eq:eq}
0=-1+\frac{\lambda_0^2\phi^2}{32\pi^2}\frac{\Lambda^4}{\bar M^2_{\rm e}(\Lambda)(\Lambda^2+\bar M^2_{\rm e}(\Lambda))^2}\,.
\eeq
Now, because $\bar M^2_{\rm e}(\Lambda)$ increases strictly with $\Lambda$,  it has a limit as $\Lambda\rightarrow 0^+$, and it has a limit or goes to $\infty$ as $\Lambda\rightarrow\infty$. From Eq.~(\ref{eq:eq}), it is easily seen that the limit as $\Lambda\rightarrow 0^+$ is necessarily $0$ and that $\bar M^2_{\rm e}(\Lambda)$ cannot diverge as $\Lambda\rightarrow\infty$. We finally obtain
\beq\label{eq:be1}
\bar M^2_{\rm e}(\Lambda)\rightarrow 0\,\,\,\,{\rm as}\,\,\,\,\Lambda\rightarrow 0^+ \quad {\rm and} \quad \bar M^2_{\rm e}(\Lambda)\rightarrow \frac{\lambda_0^2\phi^2}{32\pi^2}\,\,\,\, {\rm as}\,\,\,\,\Lambda\rightarrow\infty\,.
\eeq
We need to be more accurate concerning the behavior of $\bar M^2_{\rm e}(\Lambda)$ as $\Lambda\rightarrow 0^+$. It is easily checked that
\beq\label{eq:be2}
 \frac{\bar M^2_{\rm e}(\Lambda)}{\Lambda^2}\rightarrow\infty\,\,\,\,{\rm as}\,\,\,\,\Lambda\rightarrow 0^+\,.
 \eeq
 Similarly, $f_\Lambda(\bar M^2_{\rm e}(\Lambda))$ reads explicitly
\beq
f_\Lambda(\bar M^2_{\rm e}(\Lambda))=-\bar M^2_{\rm e}(\Lambda)+m^2-\frac{\lambda_0^2\phi^2}{32\pi^2}\left[\ln\frac{\Lambda^2+\bar M^2_{\rm e}(\Lambda)}{\bar M^2_{\rm e}(\Lambda)}-\frac{\Lambda^2}{\Lambda^2+\bar M^2_{\rm e}(\Lambda)}\right].
\eeq
Using Eqs.~(\ref{eq:be1}) and (\ref{eq:be2}), we finally obtain that
\beq
f_\Lambda(\bar M^2_{\rm e}(\Lambda))\rightarrow m^2\,\,\,\,{\rm as}\,\,\,\,\Lambda\rightarrow 0^+ \quad {\rm and} \quad f_\Lambda(\bar M^2_{\rm e}(\Lambda))\sim -\frac{\lambda_0^2\phi^2}{32\pi^2}\,\ln\Lambda^2\,\,\,\, {\rm as}\,\,\,\,\Lambda\rightarrow\infty\,.
\eeq
To summarize $f_\Lambda(\bar M^2_{\rm e}(\Lambda))$ decreases strictly from $f_{0^+}(\bar M^2_{\rm e}(0^+))=m^2>0$ to $f_\infty(\bar M^2_{\rm e}(\infty))=-\infty$. Thus there exists a critical value $\Lambda_{\rm c}$ such that $f_{\Lambda_{\rm c}}(\bar M^2_{\rm e}(\Lambda_{\rm c}))=0$ and above which $0=f_\Lambda(\bar M^2)$ has no solution, as announced earlier. Notice that what drives $f_\Lambda(\bar M^2_{\rm e}(\Lambda))$ to negative values is the presence of a logarithmic divergence. This is not a logarithmic divergence of the solution $\bar M^2$ (the later ceases to exist at $\Lambda_{\rm c}$) but a divergence of the equation itself.\\

The behavior around $\Lambda_{\rm c}$ can now be understood from the following argument. The critical cut-off $\Lambda_{\rm c}$ is defined as the particular value of $\Lambda$ at which the maximal value $f_\Lambda(\bar M^2_{\rm e}(\Lambda))$ equals zero. Denoting by $\bar M^2_{\rm c}$ the value of $\bar M^2_{\rm e}(\Lambda)$ at $\Lambda=\Lambda_{\rm c}$,\footnote{This is also the value of the (unique) solution $\bar M^2$ at $\Lambda=\Lambda_{\rm c}$.} we have $0=f'_{\Lambda_{\rm c}}(\bar M^2_{\rm c})$ and $0=f_{\Lambda_{\rm c}}(\bar M^2_{\rm c})$. If we now expand the equation $0=f_\Lambda(\bar M^2)$ around $\Lambda=\bar\Lambda_{\rm c}$ and $\bar M^2=\bar M^2_{\rm c}$, we obtain
\beq
0=f_\Lambda(\bar M^2) & = & f_{\Lambda_{\rm c}}(\bar M^2_{\rm c})+f'_{\Lambda_{\rm c}}(\bar M^2_{\rm c})(\bar M^2-\bar M^2_{\rm c})+\frac{1}{2}f''_{\Lambda_{\rm c}}(\bar M^2_{\rm c})(\bar M^2-\bar M^2_{\rm c})^2+\dots\nonumber\\
& + & \left.\frac{\partial f_\Lambda}{\partial\Lambda}\right|_{\Lambda_{\rm c},\bar M^2_{\rm c}}(\Lambda-\Lambda_{\rm c})+\dots\,,
\eeq
where in the first line we have expanded up to second order because the first two terms vanish. We then obtain that, in the vicinity of the critical point
\beq
|\bar M^2-\bar M^2_{\rm c}|\sim\left(\frac{2}{f''_{\Lambda_{\rm c}}(\bar M^2_{\rm c})}\left.\frac{\partial f}{\partial\Lambda}\right|_{\Lambda_{\rm c},\bar M^2_{\rm c}}\right)(\Lambda_{\rm c}-\Lambda)^{1/2}\,,
\eeq
which is similar to the behavior given in Eq.~(\ref{eq:behavior}).

\begin{figure}[htbp]
\begin{center}
\resizebox{0.45\textwidth}{!}{\input{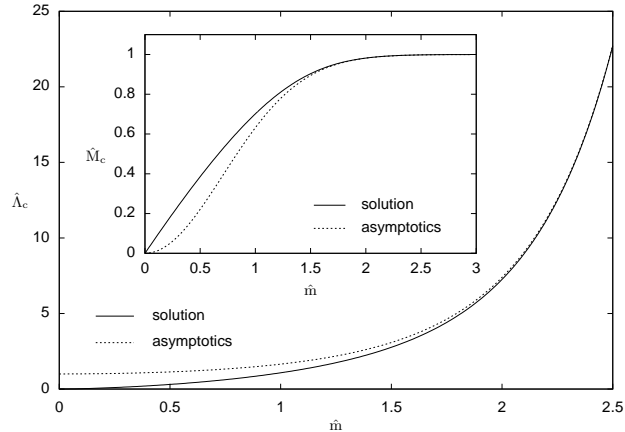}}
\caption{The dependence on the rescaled parameter $\hat m$ of the rescaled critical mass $\hat M_{\rm c}$ and the rescaled critical
cut-off $\hat\Lambda_{\rm c}$ (inset) as obtained from Eq.~(\ref{eq:asymptotic1}) in comparison to the corresponding asymptotic estimates (\ref{eq:asymptotic2}). 
\label{fig:two}}
\end{center}
\end{figure}

It is finally interesting to study how the critical cut-off depends on the parameters. It is convenient to introduce the rescaled quantities 
$\hat\phi\equiv\lambda_0 \phi/(4\pi\sqrt{2})$, $\hat m\equiv m_\phi/\hat\phi$, $\hat M_c\equiv\bar M_{\rm c}/\hat\phi,$ and $\hat \Lambda_{\rm c}\equiv\Lambda_{\rm c}/\hat\phi$, in terms of which the equations defining the critical point, that is 
$0=f'_{\Lambda_{\rm c}}(\bar M^2_{\rm c})$ and 
$0=f_{\Lambda_{\rm c}}(\bar M^2_{\rm c}),$ can be rewritten in the form
\beq
0 =  \hat m^2-\hat M^2_{\rm c}+\hat M_{\rm c}+\ln(1-\hat M_{\rm c})\,,\qquad
\hat \Lambda_{\rm c}  =  \frac{\hat M_{\rm c}^{3/2}}{(1-\hat M_{\rm c})^{1/2}}\,.
\label{eq:asymptotic1}
\eeq
A straightforward analysis then leads to
\beq\label{eq:asymptotic2}
\hat M_{\rm c}-1\sim -e^{-\hat m^2} \quad {\rm and} 
\quad \hat\Lambda_{\rm c}\sim e^{\hat m^2/2}\,,
\eeq
as $\hat m^2\rightarrow\infty$. The complete dependence of $\hat M_{\rm c}$ 
and $\hat \Lambda_{\rm c}$ with respect to $\hat m^2$ as
obtained from Eq.~(\ref{eq:asymptotic1}) is plotted in Fig.~\ref{fig:two}
and compared to the asymptotic estimates given in Eq.~(\ref{eq:asymptotic2}). We see that the asymptotic forms given in Eq.~(\ref{eq:asymptotic2}) are pretty good approximations of the exact solutions of Eq.~(\ref{eq:asymptotic1}), even for small values of $\hat m$. \\

\subsection{Combined effect of the tadpole and the bubble integral}
\begin{figure}[htbp]
\begin{center}
\resizebox{0.6\textwidth}{!}{\input{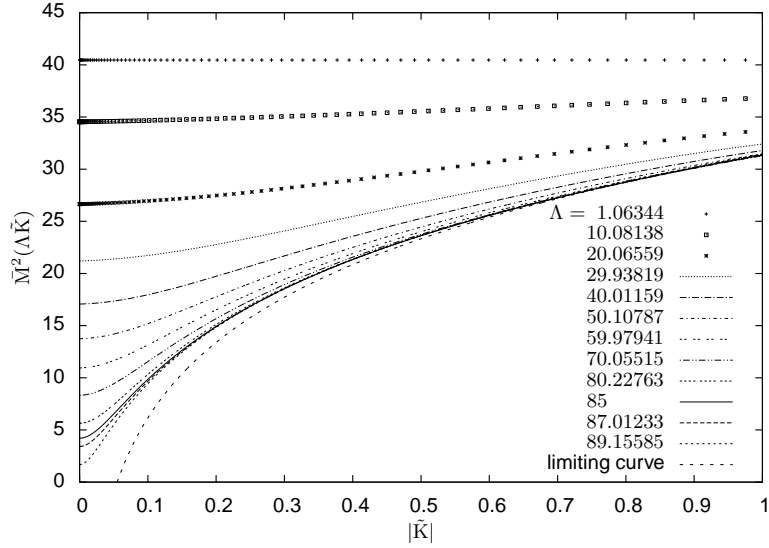}}
\caption{Evolution with $\Lambda$ of the solution $\bar M^2(K)$ of Eq.~(\ref{Eq:flow_complete}) plotted as a function of the rescaled momentum $|\tilde K|=|K|/\Lambda$. The parameters are $m^2=0.49$, $\lambda_0=20,$ and $\phi^2=4,$ while $k_{\rm m}=3.75\,10^{-4}$ and the number of discretization points is 500.
\label{fig:aaa}}
\end{center}
\end{figure}

\begin{figure}[htbp]
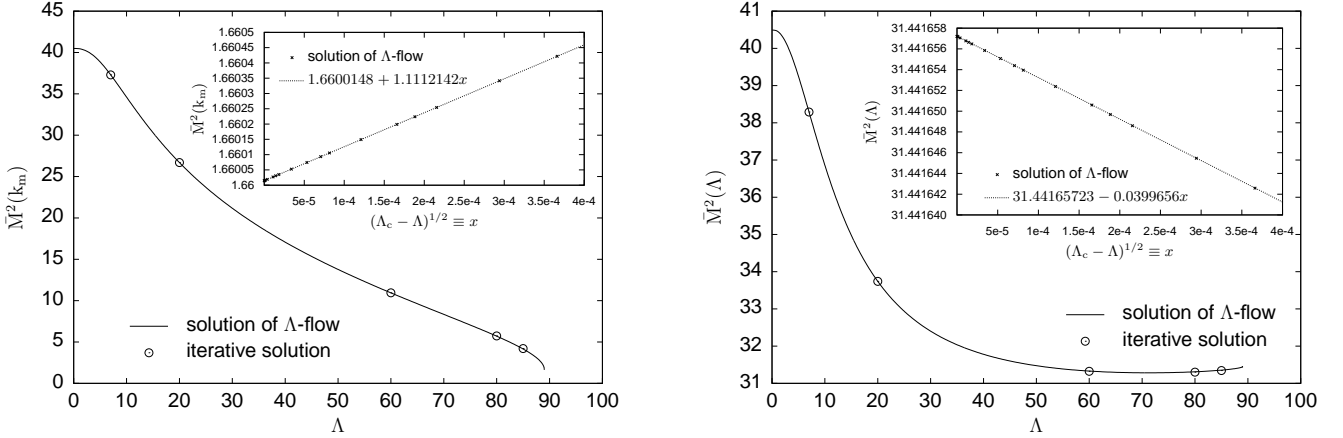

\begin{center}
\resizebox{0.45\textwidth}{!}{\input{L-flows_1st_bin.tex}}
\hspace*{1.0cm}
\resizebox{0.45\textwidth}{!}{\input{L-flows_last_bin.tex}}
\caption{Left plot: evolution with $\Lambda$ of the self-energy $\bar M^2(\Lambda\tilde K)$ evaluated for $|\tilde K|=0$ (numerically $|\tilde K|=k_{\rm m}/\Lambda$). Right plot: evolution with $\Lambda$ of the self-energy $\bar M^2(K)$ evaluated for $|\tilde K|=1$. The parameters are $m^2=0.49$, $\lambda_0=20,$ and $\phi^2=4,$ while $k_{\rm m}=7.5\,10^{-4},$ and the number of discretization points is $400$ for the solution of the flow and [$100\Lambda$] for the iterative solution.\label{fig:bb}}
\end{center}
\end{figure}

\begin{figure}[htbp]
\begin{center}
\resizebox{0.55\textwidth}{!}{\input{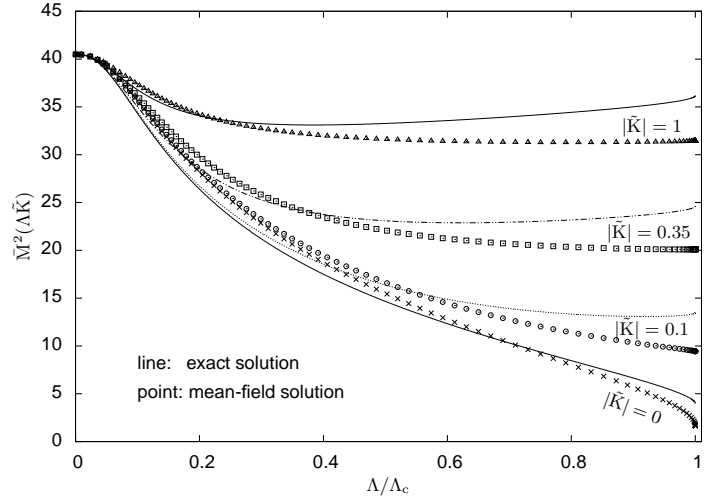}}
\caption{A comparison between the mean-field (lines) and exact (points) solution of the respective equations for $\bar M^2(\Lambda\tilde K)$ as a function of the reduced cut-off $\Lambda/\Lambda_{\rm c},$ at various values of $|\tilde K|.$ For parameters $m^2=0.49$, $\lambda_0=20$, and $\phi^2=4$, the critical cut-off is $\Lambda_{\rm c}=89.01$ in the exact and $\Lambda_{\rm c}=102.73$ 
in the mean-field case.\label{fig:d}}
\end{center}
\end{figure}

We now move to the original mass-renormalized gap equation given by Eq.~(\ref{eq:2loop2}) or Eq.~(\ref{eq:rescaled}), which includes both the effect of the tadpole and the bubble integrals. We first solve this equation for increasing values of $\Lambda$ using a flow equation that we derive along the same lines as in the previous section, see the text around Eqs.~(\ref{eq:rescaled2}) and (\ref{eq:flow}). Starting from Eq.~(\ref{eq:2loop2}) one obtains
\beq
\partial_\Lambda \bar M^2(\Lambda\tilde K)&=&\lambda_0\Lambda \int_{|\tilde Q|<1}\left[\big(\tilde Q^2+\frac{2}{\Lambda^2}\,\bar M^2(\Lambda\tilde Q)\big) \tilde G^2(\tilde Q)-\big(\tilde Q^2+\frac{2}{\Lambda^2}\, m^2\big) \tilde G_0^2(\tilde Q)\right]-\frac{\lambda_0}{2}\int_{|\tilde Q|<1}\tilde G^2(\tilde Q)
\partial_\Lambda\bar M^2(\Lambda\tilde Q)\nonumber\\
&&+\lambda_0^2\phi^2\mathop{\int_{|\tilde Q|<1}}_{|\tilde Q-\tilde K|<1}\tilde G^2(\tilde Q)\tilde G(\tilde Q-\tilde K)\left[
\frac{1}{\Lambda^2}\partial_\Lambda\bar M^2(\Lambda\tilde Q)-\frac{2}{\Lambda^3}\bar M^2(\Lambda\tilde Q)\right],
\label{Eq:flow_complete}
\eeq
where $\tilde G_0(\tilde Q)\equiv 1/(\tilde Q^2+\tilde m^2)$ and $\tilde m^2=m^2/\Lambda^2$. This equation is solved with the method outlined in App.~\ref{sec:flow} and its solution is presented in Fig.~\ref{fig:aaa}.
We observe that the behavior for increasing $\Lambda$ is similar to the one obtained previously without the inclusion of the tadpole integral and that there is again a critical value of the cut-off which cannot be overpassed. In order to save computer time, the particular values of the parameters given in the figure caption were chosen such as to observe this behavior for not too large values of $\Lambda$. A similar behavior is observed for smaller values of the coupling, but then the value of $\Lambda_{\rm c}$ is larger. An analysis of the dependence of the critical cut-off with respect to the parameters is given below. The region around $|\tilde K|=1$ is well approximated by (\ref{eq:approach}) even tough there is now an additional (momentum-independent) contribution originating from the tadpole integral. The region around $|\tilde K|=0$ is still sensitive to $\Lambda$ and, as shown in Fig.~\ref{fig:bb}, close to the critical value of the cut-off $\Lambda_{\rm c}$ at which the flow becomes singular, the solution of the flow is approximated by Eq.~(\ref{eq:behavior}). As before, this singularity propagates to any value of $\tilde K,$ and again we can try to understand the origin of the singularity using a mean-field approximation, which consists in an equation for the zero momentum self-energy
\beq\label{eq:toy2}
\bar M^2(0)\equiv \bar M^2=m^2+\frac{\lambda_0}{2}\phi^2+\frac{\lambda_0}{2}\int_{|Q|<\Lambda}\left[\frac{1}{Q^2+\bar M^2}-\frac{1}{Q^2+m^2}\right]-\frac{\lambda_0^2}{2}\,\phi^2\int_{|Q|<\Lambda}\frac{1}{(Q^2+\bar M^2)^2}\,,
\eeq
as well as a set of decoupled equations for each value $K$ (the equations are coupled to $\bar M^2$ though):
\beq\label{eq:toy4}
\bar M^2(K)=m^2+\frac{\lambda_0}{2}\phi^2+\frac{\lambda_0}{2}\int_{|Q|<\Lambda}\left[\frac{1}{Q^2+\bar M^2}-\frac{1}{Q^2+m^2}\right]-\frac{\lambda_0^2}{2}\,\phi^2\mathop{\int_{|Q|<\Lambda}}_{|Q+K|<\Lambda}\frac{1}{Q^2+\bar M^2}\frac{1}{(Q+K)^2+\bar M^2(K)}\,.
\eeq
Again, we do not expect these equations to reproduce the actual values of $\bar M^2(K)$ and $\Lambda_{\rm c}$. However, they seem to describe the large $\Lambda$ behavior correctly, at least for small values of $|\tilde K|$, see Fig.~\ref{fig:d}. Although this is not always visible, all evolutions plotted in Fig.~\ref{fig:d} present a singularity for some critical cut-off. Both in the case of the exact equation and its mean field approximation, the singularity is approached from above for small values of $|\tilde K|$ and from below for values of $|\tilde K|$ close to $1$. Note also that for $|\tilde K|\ll1$, the agreement between the exact equation and its mean field approximation is remarkably good in the whole cut-off range, once the cut-off has been rescaled by the corresponding $\Lambda_{\rm c}$. The quantitative discrepancies observed for non-small values of $|\tilde K|$ and $\Lambda$ close to $\Lambda_{\rm c}$ can be traced back to the fact that replacing the self-energy by the zero momentum self-energy in the subtracted tadpole integral is not such a good approximation as in the case of the bubble integral. Below, we will discuss another limitation of the mean-field approximation.\\

Equation (\ref{eq:toy2}) can be put in the form $0=g_\Lambda(\bar M^2)$ with
\beq\label{eq:g}
g_\Lambda(M^2)\equiv-M^2+m^2+\frac{\lambda_0}{2}\,\phi^2+\frac{\lambda_0}{32\pi^2}\left[m^2\ln\frac{\Lambda^2+m^2}{m^2}-M^2\ln\frac{\Lambda^2+M^2}{M^2}\right]-\frac{\lambda_0^2\phi^2}{32\pi^2}\left[\ln\frac{\Lambda^2+M^2}{M^2}-\frac{\Lambda^2}{\Lambda^2+M^2}\right].\ \ \ 
\eeq
It can be studied analytically along the same lines as the equation $0=f_\Lambda(\bar M^2)$ above. The shape of $g_\Lambda(M^2)$ is similar to that of $f_\Lambda(M^2)$: it increases from $g_\Lambda(0^+)=-\infty$ to $g_\Lambda(\bar M^2_{\rm e}(\Lambda))$ and then decreases to $-\infty$. The situation seems thus pretty similar to the one concerning the function $f_\Lambda(M^2)$: the existence of solutions to the equation $0=g_\Lambda(\bar M^2)$ depends on the sign of $g_\Lambda(\bar M^2_{\rm e}(\Lambda))$. However, the discussion of the sign of $g_\Lambda(\bar M^2_{\rm e}(\Lambda))$ is a little bit more involved than that of $f_\Lambda(\bar M^2_{\rm e}(\Lambda))$ and for this reason we relegate it to App.~\ref{sec:mass_ren} and focus here on the results. The main difference with respect to the case of the function $f_\Lambda(M^2)$ is that the behavior of the solution $\bar M^2$ at large $\Lambda$ depends now on the choice of parameters:\\

\begin{figure}[htbp]
\begin{center}
\includegraphics[width=8.0cm]{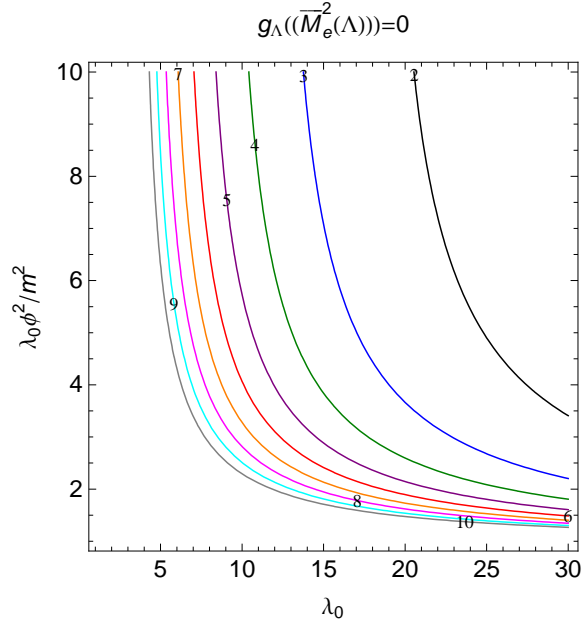}
\caption{Iso-$(\Lambda_{\rm c}/m)$ curves in the $(\lambda_0,\lambda_0\phi^2/m^2)$ plane. The label on the curves indicates the value of $\log_{10}\Lambda_c$.\label{fig:g}}
\end{center}
\end{figure}

\begin{figure}[htbp]
\begin{center}
\includegraphics[keepaspectratio,width=0.6\textwidth]{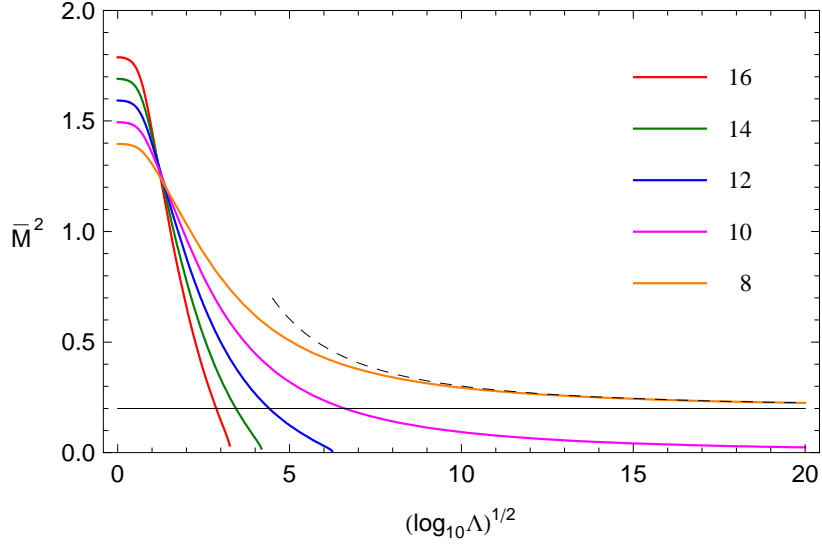}
\caption{Solution of the mean field approximated equation (\ref{eq:toy2}) for $m^2=1,$ $\phi^2=0.1$ and different values of the coupling $\lambda_0$. The figure illustrates the transition between the cases $1.$ and $2.$ discussed in the text. Starting from a situation where $m^2<\lambda_0\phi^2$, and thus such that the equation has no solution above a certain $\Lambda_{\rm c}$, the coupling is lowered down to a situation where $m^2>\lambda_0^2\phi^2$, and thus such that a continuum limit exists. As we decrease $\lambda_0$ and approach the case $m^2=\lambda_0\phi^2$, $\Lambda_{\rm c}$ increases. Right when $m^2=\lambda_0\phi^2$, $\Lambda_{\rm c}$ is still finite but incredibly large. As soon as $m^2>\lambda_0\phi^2$, there is no critical cut-off but rather a continuum limit (thin horizontal line) which is approached very slowly, as indicated by the dashed line which represents the asymptotic estimate given in Eq.~(\ref{eq:gadjodilo}). \label{fig:f}}
\end{center}
\end{figure}

$1.$ If $m^2\leq\lambda_0\phi^2$, there is a critical cut-off $\Lambda_{\rm c}$ above which the mean-field approximated equation (\ref{eq:toy2}) has no solution. This situation is similar to the one we have discussed in the previous section. The behavior in the vicinity of $\Lambda_{\rm c}$ is again $|\bar M^2-\bar M^2_{\rm c}|\propto (\Lambda_{\rm c}-\Lambda)^{1/2}$ which is also what is obtained by fitting the solution of the original equation, see Fig.~\ref{fig:bb}. This comes as no surprise since the explanation for the critical behavior is exactly the same as in the previous section, see App.~\ref{sec:mass_ren}. The plot of Fig.~\ref{fig:g} gives an idea of how $\Lambda_{\rm c}$ depends on the parameters. The critical cut-off increases rapidly as the coupling decreases, so one could argue that for moderate values of the coupling it plays no role. However, as shown in Fig.~\ref{fig:f} there is an important variation of $\bar M^2$ in the range $[0,\Lambda_{\rm c}]$ which prevents the definition of a cut-off insensitive solution. One could try to obtain a less sensitive solution by restricting the interval over which $\Lambda$ is varied. For instance, if we suppose that $m=100\,{\rm MeV}$, $\phi^2/m^2=0.1,$ and $\lambda_0=12$ then $\Lambda_{\rm c}/m\sim 10^{40},$ whereas the highest relevant scale we can conceive so far is the Plank mass, which would correspond to $\Lambda/m\sim 10^{20}$. Even if we would restrict the cut-off to vary in this window, we would observe an important variation of the mass, which prevents the definition of a cut-off independent result. If we consider as an upper cut-off the electroweak scale, $\Lambda/m\sim 10^3$, the variation is smaller but still important.\\

$2.$ If $m^2>\lambda_0\phi^2$, Eq.~(\ref{eq:toy2}) has always two-solutions. The one which is continuously connected by the flow to the unique solution at $\Lambda=0$, and which thus corresponds to the solution of the mass-renormalized gap equation plotted in Fig.~\ref{fig:aaa}, converges to the limit $m^2-\lambda_0\phi^2$ as $\Lambda\rightarrow\infty$. As we show in App.~\ref{sec:mass_ren}, this limit is approached very slowly:
\beq\label{eq:gadjodilo}
\bar M^2-(m^2-\lambda_0\phi^2)\sim \frac{1}{\ln\Lambda^2}\left((\lambda_0+48\pi^2)\phi^2+m^2\ln\frac{m^2-\lambda_0\phi^2}{m^2}\right).
\eeq
We also show that in this case, the solution $\bar M^2(K)$ of Eq.~(\ref{eq:toy4}) is defined for arbitrary large values of $\Lambda$ and that the equation is compatible with $\bar M^2(K)$ having a continuum limit $\bar M^2_\infty(K)$, solution of the following finite equation
\beq\label{eq:limit}
\bar M^2_\infty(K)=m^2-\lambda_0\phi^2-\frac{\lambda_0^2}{2}\phi^2\int_Q\frac{1}{Q^2+m^2-\lambda_0\phi^2}\left[\frac{1}{(Q-K)^2+\bar M^2_\infty(K)}-\frac{1}{Q^2+m^2-\lambda_0\phi^2}\right].
\eeq
For all practical purposes, the very slow convergence towards the continuum limit prevents the definition of cut-off independent result. In fact, this continuum limit cannot be considered too seriously because it is only reached for cut-off scales far beyond the regime of applicability of the model. As already mentioned, one could then try to define cut-off independent results by restricting the range of variation of the cut-off. But as it was already the case in the previous point, the variation of $\bar M^2$ with respect to the cut-off remains important.\\

In contrast to case 1., it is difficult to test numerically whether case 2. is relevant for the original mass renormalized equation (\ref{eq:2loop2}) simply because we cannot access such incredibly large values of the cut-off where a continuum limit could be observed. Even though case 2. could occur in certain $\Phi$-derivable approximations (see \cite{Reinosa:2011ut} for an example of truncation where it occurs), we do not believe that it occurs in the original mass-renormalized gap equation of the present $\Phi$-derivable approximation, as we now explain. Note first that the existence of a continuum limit in the mean field approximated equation can be understood as follows. Using similar manipulations as in App.~\ref{sec:subleading}, Eq.~(\ref{eq:toy2}) can be rewritten as
\beq\label{eq:ccc}
\bar M^2 & = & m^2+\frac{\lambda}{2}\phi^2+\frac{\lambda}{2}\int_{|Q|<\Lambda}(\bar M^2-m^2)^2G^2_0(Q)\bar G(Q)\nonumber\\
& - & \frac{\lambda_0^2}{2}\,\phi^2\int_{|Q|<\Lambda}\bar G^2(Q)+\frac{\lambda_0^2\lambda}{4}\phi^2\int_{|Q|<\Lambda}G^2_0(Q)\int_{|P|<\Lambda}\bar G^2(P)\,,
\eeq
with $1/\lambda=1/\lambda_0+(1/2)\int_{|Q|<\Lambda}G^2_0(Q)$. It is easy to check that this equation is compatible with the existence of a continuum limit for $\bar M^2$. Indeed, using
\beq
\int_{|Q|<\Lambda}\bar G^2(Q)=-\frac{1}{16\pi^2}\ln\frac{\Lambda^2}{m^2}+{\rm convergent}\,,
\eeq
and the fact that  $\lambda\sim 32\pi^2/\ln\Lambda^2$, one checks that the divergences in the last two contributions of Eq.~(\ref{eq:ccc}) compensate and that the only possible continuum limit is $m^2-\lambda_0\phi^2$ when $m^2>\lambda_0\phi^2$, as obtained in App.~\ref{sec:mass_ren}. The cancellation of divergences is due to the fact that the last integral in Eq.~(\ref{eq:ccc}) is the product of two decoupled single bubble integrals. It follows that its leading divergence is the square of the divergence of the bubble integral:
\beq
\int_{|Q|<\Lambda}G^2_0(Q)\int_{|P|<\Lambda}\bar G^2(P)=\frac{1}{(16\pi^2)^2}\left(\ln \frac{\Lambda^2}{m^2}\right)\left(\ln \frac{\Lambda^2}{\bar M^2}\right)+\dots\,.
\eeq
If we now use the same approach to discuss the original mass renormalized equation (\ref{eq:2loop2}), we obtain
\beq\label{eq:cccc}
\bar M^2(K) & = & m^2+\frac{\lambda}{2}\phi^2+\frac{\lambda}{2}\int_{|Q|<\Lambda}(\bar M^2(Q)-m^2)^2G^2_0(Q)\bar G(Q)\nonumber\\
& - & \frac{\lambda_0^2}{2}\,\phi^2\mathop{\int_{|Q|<\Lambda}}_{|Q-K|<\Lambda}\bar G(Q)\bar G(Q-K)+\frac{\lambda_0^2\lambda}{4}\phi^2\int_{|Q|<\Lambda}G^2_0(Q)\mathop{\int_{|P|<\Lambda}}_{|P-Q|<\Lambda}\bar G(P)\bar G(P-Q)\,,
\eeq
where we observe in particular that the momenta in the last integral are now coupled. This coupling is enough to modify the leading divergence of the integral\footnote{We assume that the self-energy does not play a role in the divergences of the bubble and the double integrals. This comes from the fact that, in the present truncation, the self-energy does not modify the leading UV contribution to the inverse propagator and also from the fact that the bubble integral has a vanishing superficial degree of divergence and that the only divergences in the double integral come from $|P|\rightarrow\infty$ or $|P|,|Q|\rightarrow\infty$, but not from $|Q|\rightarrow\infty$.} which is not anymore the square of the divergence of the bubble integral but only half of it:
\beq
\int_{|Q|<\Lambda}G^2_0(Q)\mathop{\int_{|P|<\Lambda}}_{|P-Q|<\Lambda}\bar G(P)\bar G(P-Q)=\frac{1}{2}\frac{1}{(16\pi^2)^2}\left(\ln \frac{\Lambda^2}{m^2}\right)^2+{\rm convergent}\,,
\eeq
as we show in App.~\ref{sec:evaluation}. The divergences of the r.h.s. of Eq.~(\ref{eq:cccc}) do not cancel anymore and the right-hand-side behaves as $-(\lambda_0^2\phi^2/(32\pi^2)^2)(\ln\Lambda^2/m^2)^2$. This is incompatible with the existence of a continuum limit since the solution of the gap equation cannot become negative. It is then most probable that the solution of the original mass-renormalized gap equation ceases to exist beyond some value of the cut-off, just as described in case 1. Anyway, irrespectively of the fact that case 2. is relevant for the original mass renormalized equation or not, it is clear from our analysis that renormalizing the mass without renormalizing the coupling, although it does not lead to a divergence of the solution of the mass equation, is not sufficient to ensure the insensitivity of the solution with respect to the cut-off. In the next section, we cure this problem by means of coupling renormalization.\\

\section{Completely renormalized equation}\label{sec:three}
The analysis of the previous section shows that the presence of logarithmic divergences in the mass-renormalized gap equation does not always translate into logarithmic divergences of its solution. Instead, these divergences are responsible for the appearance of a critical cut-off $\Lambda_{\rm c}$ above which the gap equation has no more solutions or could also lead to the existence of a continuum limit which is however only reached for incredibly large, and practically inaccessible, values of the cut-off. These features prevent the existence of solutions of the gap equation which are insensitive to the cut-off. In this section, we explain how to get rid of these effects and define a {\it completely renormalized gap equation} whose results are insensitive to the cut-off already for moderate values of the cut-off (up to terms of order $1/\Lambda$ when one uses a sharp cut-off). The idea is to use coupling renormalization, not to absorb subdivergences of the self-energy $\bar M^2(K)$ (there are no such divergences as we have seen), but to eliminate the remaining divergences of the gap equation. In this way, one can hope to get rid of the undesirable effects mentioned above.\\

\subsection{Renormalization method}
The main difficulty is that one should be able to perform the renormalization program outlined above using a field-independent bare coupling. It is not necessarily possible to do so if one keeps the original form of the gap equation. We know how to proceed if we slightly modify the original equation into \cite{Berges:2005hc}
\beq\label{eq:mgap}
\bar M^2(K) & = & m^2+\frac{\lambda_2}{2}\,\phi^2+\frac{\lambda_0}{2}\int_{|Q|<\Lambda}\Big[\bar G(Q)-G_0(Q)\Big]-\frac{\lambda^2}{2}\,\phi^2\mathop{\int_{|Q|<\Lambda}}_{|Q-K|<\Lambda}\bar G(Q)\bar G(Q-K)\,,
\eeq
by allowing the bare couplings $\lambda_0$ and $\lambda_2$ to be different.\footnote{These two different bare couplings should be seen as two different approximations to the unique bare coupling of the exact theory. The reason why these two bare couplings are different in a given truncation is that they will renormalize divergences originating from different diagrammatic topologies.} Note also that the bubble integral is multiplied by $\lambda$ which will later become the renormalized coupling. There exist different but equivalent approaches to explain how the bare couplings $\lambda_0$ and $\lambda_2$ need to be chosen. Here we follow a method which, although it does not allow to capture the general structure behind the renormalization of $\Phi$-derivable approximations, has the advantage of being very similar to the approach used to renormalize the gap equation in the presence of the tadpole integral only.\footnote{In a forthcoming work, we shall apply the same method at finite temperature. There, we will make contact with the more general approaches developed in \cite{Berges:2005hc}.}\\

The first step is to decompose the self-energy into a {\it local} and a {\it non-local} part $\bar M^2(K)=\bar M^2_{\rm l}+\bar M^2_{\rm nl}(K)$ with
\beq
\bar M^2_{\rm l} & = & m^2+\frac{\lambda_{2,{\rm l}}}{2}\,\phi^2+\frac{\lambda_0}{2}\int_{|Q|<\Lambda}\Big[\bar G(Q)-G_0(Q)\Big]\,,\\
\bar M^2_{\rm nl}(K) & = & \frac{\lambda_{2,{\rm nl}}}{2}\,\phi^2-\frac{\lambda^2}{2}\,\phi^2\mathop{\int_{|Q|<\Lambda}}_{|Q-K|<\Lambda}\bar G(Q)\,\bar G(Q-K)\,,\label{eq:th}
\eeq
where, for convenience, we have split the bare parameter $\lambda_2$ as $\lambda_2=\lambda_{2,{\rm l}}+\lambda_{2,{\rm nl}}$. The decomposition of the self-energy into local and non-local parts was used also in \cite{Reinosa:2003qa,Arrizabalaga:2006hj,Fejos:2009dm} and it naturally arises when working in the auxiliary field formalism. The integral in Eq.~(\ref{eq:th}) introduces a cut-off sensitivity in the gap equation which, as we have seen, does not lead to a divergence of the solution but has some undesirable effects. To get rid of those, we can try to absorb the sensitivity by choosing
\beq\label{eq:dl2nl}
\lambda_{2,{\rm nl}}=\lambda^2\int_{|Q|<\Lambda}G^2_0(Q)\,.
\eeq
The usual consensus is that this choice is enough to absorb the sensitivity of $\bar M^2_{\rm nl}(K)$ with respect to the cut-off because we expect that this sensitivity does not depend on the self-energy. In fact the self-energy is expected to grow logarithmically at large momentum. \\

Let us now treat the local contribution $\bar M^2_{\rm l}$. Using the identity:
\beq
\bar G(Q)=G_0-(\bar M^2(Q)-m^2)G_0(Q)\bar G(Q)=G_0-(\bar M^2(Q)-m^2)G^2_0(Q)+(\bar M^2(Q)-m^2)^2G^2_0(Q)\bar G(Q)\,,
\eeq
 we write $\bar M^2_{\rm l}$ as
\beq
\bar M^2_{\rm l} & = & m^2+\frac{\lambda_{2,{\rm l}}}{2}\,\phi^2-\frac{\lambda_0}{2}\int_{|Q|<\Lambda}(\bar M^2(Q)-m^2)G^2_0(Q)+\frac{\lambda_0}{2}\int_{|Q|<\Lambda}(\bar M^2(Q)-m^2)^2G^2_0(Q)\bar G(Q)\,.
\eeq
We expect the last integral not to be sensitive to large values of the cut-off. In contrast, the first integral is sensitive to $\Lambda.$ Using the decomposition of $\bar M^2(Q)$ into a local and a non-local part in the first integral, we arrive at
\beq\label{eq:aa}
(\bar M^2_{\rm l}-m^2)\left[\frac{1}{\lambda_0}+\frac{1}{2}\int_{|Q|<\Lambda}G^2_0(Q)\right] & = & \frac{\lambda_{2,{\rm l}}}{\lambda_0}\frac{\phi^2}{2}-\frac{1}{2}\int_{|Q|<\Lambda}\bar M^2_{\rm nl}(Q)G^2_0(Q)\nonumber\\
& + & \frac{1}{2}\int_{|Q|<\Lambda}(\bar M^2(Q)-m^2)^2G^2_0(Q)\bar G(Q)\,.
\eeq
We are now in a situation where we can use the same trick as in the case where the tadpole contribution is present only. We set
\beq\label{eq:dl0}
\frac{1}{\lambda_0}=\frac{1}{\lambda}-\frac{1}{2}\int_{|Q|<\Lambda}G^2_0(Q)\,,
\eeq
which leads to
\beq\label{eq:th2}
\bar M^2_{\rm l}=m^2+\frac{\lambda}{2}\left[\frac{\lambda_{2,{\rm l}}}{\lambda_0}\,\phi^2-\int_{|Q|<\Lambda}\bar M^2_{\rm nl}(Q)G^2_0(Q)\right]+\frac{\lambda}{2}\int_{|Q|<\Lambda}(\bar M^2(Q)-m^2)^2G^2_0(Q)\bar G(Q)\,.
\eeq
However, the first integral is still potentially sensitive to the cut-off. We could absorb this sensitivity by adjusting $\lambda_{2,{\rm l}}$ but for this to be possible the sensitivity should be proportional to $\phi^2$. To check this, note that $\bar M^2_{\rm nl}(K)$ can be written as
\beq\label{eq:bb}
\bar M^2_{\rm nl}(K) & = & \frac{\lambda^2}{2}\,\phi^2\left[\int_{|Q|<\Lambda}G^2_0(Q)-\mathop{\int_{|Q|<\Lambda}}_{|Q-K|<\Lambda}G_0(Q)G_0(Q-K)\right]\nonumber\\
& + & \frac{\lambda^2}{2}\,\phi^2\left[\mathop{\int_{|Q|<\Lambda}}_{|Q-K|<\Lambda}G_0(Q)G_0(Q-K)-\mathop{\int_{|Q|<\Lambda}}_{|Q-K|<\Lambda}\bar G(Q) \bar G(Q-K)\right].
\eeq
Because we expect the second term of this equation to decrease fast enough at large $|K|$, the sensitivity of the first integral in Eq.~(\ref{eq:th2}) originates exclusively from the first line of Eq.~(\ref{eq:bb}) and is thus proportional to $\phi^2$, as needed. The remaining sensitivity in Eq.~(\ref{eq:th2}) can thus be absorbed by adjusting $\lambda_{2,{\rm l}}$ such that
\beq\label{eq:dl2}
\frac{\lambda_{2,{\rm l}}}{\lambda_0}-\frac{\lambda^2}{2}\int_{|Q|<\Lambda}G_0^2(Q)\left[\int_{|R|<\Lambda}G^2_0(R)-\mathop{\int_{|R|<\Lambda}}_{|R-Q|<\Lambda}G_0(R)G_0(R-Q)\right]=1\,.
\eeq
The choice of $1$ in the r.h.s. of this condition is such that $\lambda_2=\lambda_0+O(\lambda^2)$, as it should be at this order of accuracy.\footnote{We mention here that the splitting of $\lambda_2$ in local and non-local parts is completely arbitrary and that the actual expression of both parts depend on the procedure used to determine them. Indeed, the interested reader can check that if one multiplies Eq.~(\ref{eq:aa}) by $\lambda_0$ and writes its l.h.s. in a form similar to that in (\ref{eq:sept}), then the inversion of the resulting operator (see (\ref{eq:inversion})) results in a different form for $\delta\lambda_{2,{\rm l}}$ and $\delta\lambda_{2,{\rm nl}}$. But this is not a problem at all, for what matters is that this different procedure gives the very same expression for $\lambda_2$ in terms of divergent integrals.} Using  Eq.~(\ref{eq:dl2}) and (\ref{eq:bb}) in Eq.~(\ref{eq:th2}) and adding the non-local contribution (\ref{eq:th}) with the choice of $\lambda_{2,{\rm nl}}$ given in Eq.~(\ref{eq:dl2nl}), we finally arrive at the completely renormalized gap equation
\beq\label{eq:complete}
\bar M^2(K) & = & m^2+\frac{\lambda}{2}\,\phi^2+\frac{\lambda}{2}\int_{|Q|<\Lambda}(\bar M^2(Q)-m^2)^2G^2_0(Q)\bar G(Q)\nonumber\\
& - & \frac{\lambda^2}{2}\,\phi^2\left[\mathop{\int_{|Q|<\Lambda}}_{|Q-K|<\Lambda}\bar G(Q) \bar G(Q-K)-\int_{|Q|<\Lambda}G^2_0(Q)\right]\nonumber\\
& + & \frac{\lambda^3}{4}\phi^2\int_{|Q|<\Lambda}G^2_0(Q)\left[\mathop{\int_{|R|<\Lambda}}_{|R-Q|<\Lambda}\bar G(R)\bar G(R-Q)-\mathop{\int_{|R|<\Lambda}}_{|R-Q|<\Lambda}G_0(R)G_0(R-Q)\right].
\eeq
An equation of similar form was derived using slightly different renormalization conditions in \cite{Arrizabalaga:2006hj} and also in \cite{Fejos:2011zq}. It was solved together with the field equation at finite temperature in Minkowski space \cite{Arrizabalaga:2006hj} and at zero temperature in Euclidean space \cite{Fejos:2011zq}. Here, we solve this equation for a fixed value of $\phi$. 

\subsection{Numerical solution of Eq.~(\ref{eq:complete}) and discussion}

\begin{figure}[t]
\begin{center}
\resizebox{0.6\textwidth}{!}{\input{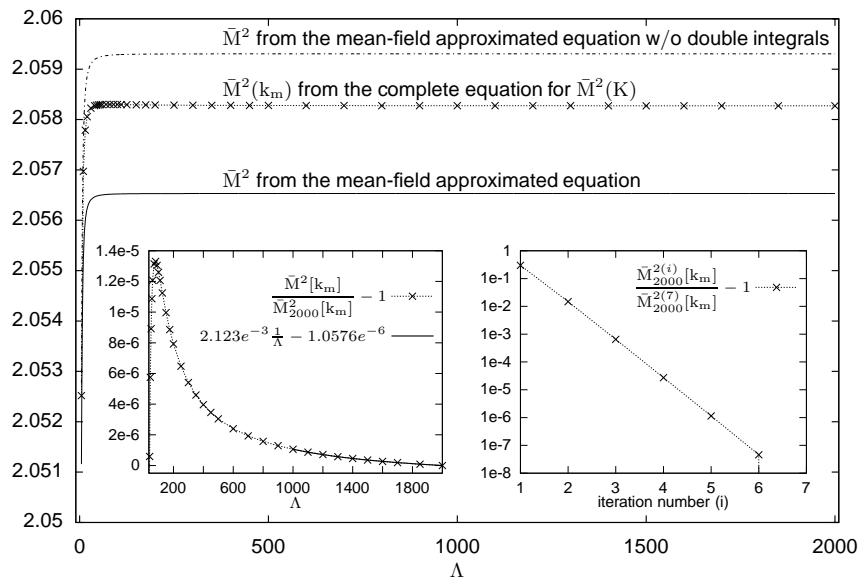}}
\caption{
The cut-off dependence of the converged value of $\bar M^2(k_{\rm m})$ obtained from the solution of the completely renormalized equation as compared to the solution $\bar M^2$ of the toy equation with and without the inclusion of the two double integrals.  $\bar M^2(k_{\rm m})$ converged with $10^{-7}$ accuracy after 7 iterations. The inset at the left shows that $\bar M^2(k_{\rm m})$ depends non monotonously on $\Lambda$ and approaches the asymptotic value as $1/\Lambda.$ The inset at the right shows the approach of the converged value of $\bar M^2(k_{\rm m})$ in the case of the highest cutoff used. The model parameters are: $m^2=1,$ $\lambda=10,$ and $\phi^2=0.2.$ The number of discretization points is [100$\Lambda$] and $k_{\rm m}=5.10^{-4}.$
\label{fig:finite}}
\end{center}
\end{figure}

Figure~\ref{fig:finite} represents the numerical solution of Eq.~(\ref{eq:complete}) for increasing values of $\Lambda$ and of some approximated versions of this equation obtained by using a mean-field approximation and keeping or dropping the double integrals in Eq.~(\ref{eq:complete}). The solution of these equations are compatible with the existence of a continuum limit, which is approached relatively fast, as $1/\Lambda$ in the case of the sharp cut-off that we have chosen here. It is important to realize that the bare coupling redefinitions that we have obtained are precisely such that, if one would expand the solution of the original bare gap equation (\ref{eq:mgap}) in powers of the renormalized coupling $\lambda$, the coefficients of the expansion would converge as $\Lambda\rightarrow\infty$, see \cite{Fejos:2011zq} for a diagrammatic representation of this fact. Thus, even though the divergences of such a perturbative expansion do not always appear as divergences of the solution of the gap equation, the ``correct'' renormalization procedure needs to get rid of those as well, in order to avoid undesirable features such as those we have pointed out previously. This reinforces our belief that, although the two-point function is defined non-perturbatively, constructing a renormalization procedure which remove those divergences which appear when expanding formally the two-point function in powers of the coupling, is a sensible thing to do. We insist once again that this is a priori not obvious because the object we are finally considering is the solution of a self-consistent equation, not its perturbative expansion.\\

Note that without a deeper analysis of the existence and the nature of the solutions of Eq.~(\ref{eq:complete}), we cannot really conclude to the existence of a continuum limit. But this is not so important, for what really matters is that there exists a wide range of cut-off scales, far above $m$ and $\phi$, where the solution exists and is almost insensitive to the cut-off, to within power law corrections, as it is shown in Fig.~\ref{fig:finite}. In fact from Eq.~(\ref{eq:dl0}) it follows that, if one wants to maintain $\lambda_0>0$, one has to choose $\lambda>0$ and $\Lambda$ below a certain scale $\Lambda_{\rm p}$ (Landau scale) defined by
\beq\label{eq:Lp}
0=\frac{1}{\lambda}-\frac{1}{2}\int_{|Q|<\Lambda_{\rm p}}\frac{1}{(Q^2+m^2)^2}\approx \frac{1}{\lambda}-\frac{1}{32\pi^2}\ln\frac{\Lambda^2_{\rm p}}{m^2e}\,.
\eeq
In the regime of interest, that is when $m,\phi\ll\Lambda,\Lambda_{\rm p}$, we can fulfill simultaneously the requirements that the solution is almost insensitive to the cut-off $\Lambda$ and that $\Lambda<\Lambda_{\rm p}$. Taking values of $\Lambda$ above $\Lambda_{\rm p}$ almost does not change the solution, although this corresponds to negative $\lambda_0$. Note also that, as long as $\lambda_0>0$, we have $\lambda_2>0$. This is because $\lambda_{2,{\rm nl}}>0$ and 
\beq
\lambda_{2,{\rm l}}=\lambda_0\left[1+\frac{\lambda^2}{2}\int_{|Q|<\Lambda}G_0^2(Q)\Big[B_0(Q)-B_0(0)\Big]\right]
\eeq
is positive since, as we show in Appendix~\ref{sec:evaluation}, the perturbative bubble integral defined with $G_0$ as
\beq
B_0(K)\equiv -\mathop{\int_{|Q|<\Lambda}}_{|Q-K|<\Lambda} G_0(Q) G_0(Q-K),
\eeq
is a monotonously increasing function of $|K|.$\\

Another interesting question is the following. We have considered a renormalization condition for the mass at $\phi^2=0$ which somehow assumes that the system is in the symmetric phase (since $\phi=0$ is accessible). Because the self-energy at zero momentum should represent the curvature of the effective potential (in the exact theory at least), we expect it to be defined and positive for any value of $\phi$ (at least as long as $\phi$ does not become of the order of the Landau scale $\Lambda_{\rm p}$). However this is not necessarily guaranteed because the completely renormalized gap equation involves contributions with different signs. This issue will also appear when studying the problem at finite temperature because we shall perform renormalization at some high temperature $T_\star$ where we will require that the system is in its symmetric phase by imposing that the self-energy at zero momentum and zero field be positive. We will then have to check, probably numerically, that the self-energy at $T=T_\star$ and zero momentum remains positive as we increase the field. Here, we propose a discussion of this issue, at $T=0$ and on a simplified version of the gap equation which allows for certain analytical arguments. Our main purpose is to show that $\bar M^2(0)$ should remain defined in some range of parameters, which we name the regime of interest, that is when $m,\phi\ll\Lambda,\Lambda_{\rm p}$. We consider the equation 
\beq
\bar M^2 & = & m^2+\frac{\lambda}{2}\,\phi^2+\frac{\lambda}{2}\int_{|Q|<\Lambda}\left[\frac{1}{Q^2+\bar M^2}-\frac{1}{Q^2+m^2}+\frac{\bar M^2-m^2}{(Q^2+m^2)^2}\right]\nonumber\\
& - & \frac{\lambda^2}{2}\,\phi^2\int_{|Q|<\Lambda}\left[\frac{1}{(Q^2+\bar M^2)^2}-\frac{1}{(Q^2+m^2)^2}\right]\,,
\label{eq:toy_wo_double}
\eeq
obtained from Eq.~(\ref{eq:complete}) by performing a mean-field approximation, as in previous sections, for the zero momentum self-energy and by dropping the last line which contains double integrals. The solution of the completely renormalized gap equation in the mean-field approximation can be seen in Fig.~\ref{fig:finite}, both without and with the inclusion of the double integrals. Interestingly, the solution obtained without the inclusion of the double integrals, that is of Eq.~(\ref{eq:toy_wo_double}), is closer to the solution of the original equation (\ref{eq:complete}). For this reason we find it worthwhile to study this simplified version of the equation. The latter can be put in the form $0=h_\Lambda(\bar M^2),$ so that for the first and second derivatives one has
\beq
h'_\Lambda(M^2) & = & -1-\frac{\lambda}{2}\int_{|Q|<\Lambda}\frac{1}{(Q^2+M^2)^2}+\frac{\lambda}{2}\int_{|Q|<\Lambda}\frac{1}{(Q^2+m^2)^2}+\lambda^2\phi^2\int_{|Q|<\Lambda}\frac{1}{(Q^2+M^2)^3}\,,\\
h''_\Lambda(M^2) & = & \lambda\int_{|Q|<\Lambda}\frac{1}{(Q^2+M^2)^3}-3\lambda^2\phi^2\int_{|Q|<\Lambda}\frac{1}{(Q^2+M^2)^4}\,.
\eeq
The second derivative of $h_\Lambda(M^2)$ is identical to that of $g_\Lambda(M^2)$ which we discuss in App.~\ref{sec:mass_ren} with the only exception that $\lambda_0$ is now replaced by $\lambda$. Because $\lambda>0$, we conclude then that $h''_\Lambda(M^2)$ vanishes only once as $M^2$ varies from $0$ to $\infty$. It is negative for small enough $M^2$, and positive for large enough $M^2$. It follows that $h'_\Lambda(M^2)$ first decreases from $h'_\Lambda(0)=\infty$ and then increases towards 
\beq
h'_\Lambda(\infty)=-1+\frac{\lambda}{2}\int_{|Q|<\Lambda}\frac{1}{(Q^2+m^2)^2}\,.
\eeq
The rest of the discussion depends on the sign of $h'_\Lambda(\infty)$ and thus on the location of $\Lambda$ with respect to the Landau scale $\Lambda_{\rm p}$:\\

1. If $\Lambda<\Lambda_{\rm p}$, $h'_\Lambda(\infty)<0$. Then $h'_\Lambda(M^2)$ vanishes only once for $M^2=\bar M^2_{\rm e}(\Lambda)$ and $h_\Lambda(M^2)$ increases from $h_\Lambda(0)=-\infty$ to $h_\Lambda(\bar M^2_{\rm e}(\Lambda))$ and then decreases towards $h_\Lambda(\infty)=-\infty$. This is a situation that we have already encountered: the number of solutions depends on the sign of $h_\Lambda(\bar M^2_{\rm e}(\Lambda))$. Now, it is easily checked that $h_\Lambda(m^2)=\lambda\phi^2/2$ and thus $h_\Lambda(\bar M^2_{\rm e}(\Lambda))\geq 0$. Then, as long as $\Lambda<\Lambda_{\rm p}$, there are two solutions (which can be degenerate). We will be more particularly interested in the right most solution, which we name the ``physical solution'', for it is continuously connected to the unique solution for $\Lambda=0$ and thus corresponds to the solution of the completely renormalized gap equation that we have plotted in Fig.~\ref{fig:finite}.\\

2. If $\Lambda=\Lambda_{\rm p}$, $h'_\Lambda(\infty)=0$. We can draw the same conclusions as above with the only exception that, for large enough $M^2$, $h_\Lambda(M^2)$ does not decrease towards $-\infty$ but towards
\beq
h_\Lambda(\infty)=\frac{\lambda}{2}\phi^2\left[1+\lambda\int_{|Q|<\Lambda_{\rm p}}\frac{1}{(Q^2+m^2)^2}\right]-\frac{\lambda}{2}\int_{|Q|<\Lambda_{\rm p}}\frac{1}{Q^2+m^2}=\frac{3\lambda}{2}\phi^2-\frac{\lambda}{2}\int_{|Q|<\Lambda_{\rm p}}\frac{1}{Q^2+m^2}\,.
\eeq
Then, if $\phi^2\geq\phi^2_{\rm p}$, with
\beq
\phi^2_{\rm p}\equiv \frac{1}{3}\int_{\Lambda_{\rm p}}\frac{1}{Q^2+m^2}\,,
\eeq
the physical solution disappears. However in the regime of interest, such that in particular $\Lambda_{\rm p}\gg m$, this situation corresponds to extremely large values of the field, namely $\phi^2\geq\Lambda^2_{\rm p}/48\pi^2$. In contrast, if $\phi^2<\phi^2_{\rm p}$, the physical solution subsists.\\

3. If $\Lambda>\Lambda_{\rm p}$, $h'_\Lambda(\infty)>0$. There are three cases depending on the sign of the minimal value reached by $h'_\Lambda(M^2)$. Without trying to be exhaustive let us analyze what happens in the limit $\Lambda\rightarrow\infty$. We have
\beq
h'_\infty(M^2)=-1+\frac{\lambda}{32\pi^2}\left[\ln\frac{M^2}{m^2}+\frac{\lambda\phi^2}{M^2}\right].
\eeq
It is easily seen that $h'_\infty(M^2)$ reaches its minimal value for $M^2=\lambda\phi^2$, that is
\beq\label{eq:hp}
h'_\infty(\lambda\phi^2)=-1+\frac{\lambda}{32\pi^2}\left[\ln\frac{\lambda\phi^2}{m^2}+1\right].
\eeq
If this quantity is positive, $h'_\infty(M^2)>0$ and $h_\infty(M^2)$ increases from $-\infty$ to $\infty$. There is one solution but it does not correspond to the physical solution. However, comparing (\ref{eq:hp}) and (\ref{eq:Lp}), we observe that this situation occurs only if $\lambda\phi^2e^2\geq\Lambda_{\rm p}^2$, that is for extremely large values of the field. In the opposite case, $h'_\infty(M^2)$ vanishes twice, for $M^2=\bar M^2_{\rm e}(\infty)$ and $M^2=\bar M^2_{\rm f}(\infty)$ and thus $h_\infty(M^2)$ increases from $-\infty$ to $h_\infty(\bar M^2_{\rm e}(\infty))$, then decreases to $h_\infty(\bar M^2_{\rm f}(\infty))$ and finally increases again towards $\infty$. There can be up to three solutions, the physical one corresponding to the ``second'' one. For the later to exist, we must have $h_\infty(\bar M^2_{\rm f}(\infty))<0$. 
We have
\beq
h_\infty(\bar M^2_{\rm f}(\infty))=-\bar M^2_{\rm f}(\infty)+m^2+\frac{\lambda}{2}\phi^2+\frac{\lambda}{32\pi^2}\left[(\bar M^2_{\rm f}(\infty)+\lambda\phi^2)\ln\frac{\bar M^2_{\rm f}(\infty)}{m^2}-\bar M^2_{\rm f}(\infty)+m^2\right],
\eeq
with $\bar M^2_{\rm f}(\infty)$ the right-most solution of
\beq
0=-1+\frac{\lambda}{32\pi^2}\left[\ln\frac{\bar M^2_{\rm f}(\infty)}{m^2}+\frac{\lambda\phi^2}{\bar M^2_{\rm f}(\infty)}\right].
\eeq
Then, we can also write
\beq
h_\infty(\bar M^2_{\rm f}(\infty)) & = & m^2\left(1+\frac{\lambda}{32\pi^2}\right)+\lambda\phi^2\left(\frac{3}{2}-\frac{\lambda^2}{32\pi^2}\frac{\phi^2}{\bar M^2_{\rm f}(\infty)}\right)
-\frac{\lambda\bar M^2_{\rm f}(\infty)}{32\pi^2}\left(1+\frac{\lambda\phi^2}{\bar M^2_{\rm f}(\infty)}\right).
\eeq
Now, it is easy to check that $\bar M^2_{\rm f}(\infty)$ decreases as $\phi^2$ increases, from $\bar M^2_{\rm f}(\infty)\sim \Lambda^2_{\rm p}/e$ when $\phi^2\rightarrow 0$ to $0$ as $\phi^2$ approaches $\Lambda^2_{\rm p}/\lambda e^2$. Thus, in the regime of interest, $\bar M^2_{\rm f}(\infty)$ is way larger than $m^2$ and $\phi^2$ and $h_\infty(\bar M^2_{\rm f}(\infty))<0$, as announced. To summarize, this indicates that in the regime of interest the physical solution can be continued from $\Lambda<\Lambda_{\rm p}$ to arbitrary large values of $\Lambda$.

\section{Concluding remarks}
Using a particular example of truncation where the two-point function has a non-trivial momentum and field dependence, we have revisited the renormalization of $\Phi$-derivable approximations paying particular attention to the question of the existence of a solution of the gap equation for arbitrarily large values of the cut-off and to the fact that some of the perturbative divergences which appear when expanding the solution of the gap equation in powers of the coupling do not appear as divergences at the level of the non-expanded solution. We have shown that it was nevertheless important to absorb these ``perturbative'' divergences in order to avoid certain inconvenient features.\\ 

Our analysis contributes to clarify the meaning of the renormalization procedure for $\Phi$-derivable approximations which has been constructed in recent years. It shows in particular that the formal perturbative expansions which have been used sometimes to construct a renormalization scheme for $\Phi$-derivable approximations are a good guiding principle, although the corresponding perturbative divergences not always appear as divergences at the level of the solution of the gap equation. It also points to the fact that approximated renormalization schemes where (at least the leading) logarithmic sensitivities are not completely eliminated need to be considered with care. All these remarks go beyond the particular framework of $\Phi$-derivable approximations and probably apply to other resummation methods such as Schwinger-Dyson equations.\\

Finally this work is an important step towards the inclusion of finite temperature effects. In fact, once the renormalization has been properly performed at zero temperature, the equations at finite temperature should be automatically renormalized. We plan to study this truncation at finite temperature in the imaginary time formalism, extending our approach in \cite{Reinosa:2011ut} and investigate whether or not it leads to a correct order for the phase transition, as it was already claimed in \cite{Arrizabalaga:2006hj} from numerical results obtained in the real-time formalism.

\acknowledgments{We would like to thank Julien Serreau for useful comments. We also thank Gergely Mark{\'o} for discussions on some of the numerical aspects of this work.}

\appendix

\section{General regularization for the 2PI effective action}\label{sec:reg}
Consider the following generating functional
\beq\label{eq:Z}
Z[J,K]\equiv\frac{\int {\cal D}\varphi\,\exp\,\left\{-\frac{1}{2}\,\varphi\cdot (G_0R)^{-1}\cdot\varphi+S_{\rm int}[\varphi]+J\cdot\varphi+\frac{1}{2}\,\varphi\cdot K\cdot\varphi\right\}}{\int {\cal D}\varphi\,\exp\,\left\{-\frac{1}{2}\,\varphi\cdot (G_0R)^{-1}\cdot\varphi\right\}}\,,
\eeq
where $J\cdot\varphi\equiv\int_x J(x)\varphi(x)$, $\varphi\cdot K\cdot\varphi\equiv\int_x\int_y\varphi(x)K(x,y)\varphi(y)$ and $\varphi\cdot(G_0R)^{-1}\cdot\varphi\equiv\int_x\int_y\int_z\varphi(x)G^{-1}_0(x,y)R^{-1}(y,z)\varphi(z)$. For $K=0$, the functional $Z[J,0]$ is regularized due to the presence of the regulator $R$ and the normalization factor $1/\int{\cal D}\varphi\,\exp\{-\frac{1}{2}\,\varphi\cdot (G_0R)^{-1}\cdot\varphi\}$. To see this explicitly, one can consider a perturbative expansion which turns the evaluation of $Z[J,0]$ into the evaluation of Gaussian integrals. Wick theorem for Gaussian integrals generates an overall infinite determinant factor which is precisely canceled by the normalization factor. If follows an expansion of $Z[J,0]$ organized in terms of Feynman integrals involving the propagator $G_0R$ and thus properly regularized for an appropriate choice of $R$. In what follows, we will only consider the functional $Z[J,K]$ in the vicinity of $K=0$. For those ``small'' values of $K$, we expect the regularization at $K=0$ to be sufficient to regularize $Z[J,K]$ as well.\\

The 2PI effective action $\Gamma[\phi,G]$ can be seen as the double Legendre transform of $Z[J,K]$ with respect to the sources $J$ and $K$. Taking properly into account the presence of a regulator $R$ and a normalization factor, one obtains
\beq\label{eq:2PI}
\Gamma[\phi,G]=\frac{1}{2}\,\phi\cdot(G_0R)^{-1}\cdot\phi+\frac{1}{2}{\rm Tr}\,\left[\ln G^{-1}-\ln (G_0R)^{-1}+(G_0R)^{-1}\cdot G-1\right]+\Gamma_{\rm int}[\phi,G]\,,
\eeq
where the first two terms can be obtained by switching off all interactions in Eq.~(\ref{eq:Z}) and $\Gamma_{\rm int}[\phi,G]$ includes all contributions due to interactions in the form of two-particle irreducible diagrams with propagator $G$. The full propagator $\bar G[\phi]$ corresponding to the functional $Z[J,K=0]$ can be obtained from a variational principle applied to $\Gamma[\phi,G]$ that is
\beq
0=\left.\frac{\delta\Gamma}{\delta G}\right|_{\phi,\bar G[\phi]}\,.
\eeq
This condition can be written equivalently as
\beq\label{eq:2PIgap}
\bar G^{-1}[\phi]=(G_0R)^{-1}+2\!\left.\frac{\delta\Gamma_{\rm int}}{\delta G}\right|_{\phi,\bar G[\phi]}\,.
\eeq
Note that it is a priori not obvious that the expressions (\ref{eq:2PI}) and (\ref{eq:2PIgap}) are regularized, specially because the regulator $R$ does not appear explicitly in the Feynman integrals contributing to $\Gamma_{\rm int}[\phi,G]$. In what follows, we address the question of how the regularization of $Z[J,K]$ appears at the level of $\Gamma[\phi,G]$. We do so in a simple situation where the source $K(x,y)$ is translation invariant, allowing us to conveniently work in Fourier space.\\

It is important to point out that the regularization of $Z[J,K]$ is a priori effective when $K$ is small enough. In terms of the conjugated variables this means that $G$ is close to $\bar G$. The later has a UV asymptotic behavior $\sim G_0R$, because except from the momentum independent contributions originating from the tadpole integral, the quantum fluctuations that contribute to $\bar G$ are suppressed in the UV by the regulator $R$. It follows that in the vicinity of $\bar G$, which is the only region where we need to consider the functional $\Gamma[\phi,G]$, the Feynman integrals contributing to $\Gamma_{\rm int}[\phi,G]$ are all regularized. It remains to discuss the trace contribution of Eq.~(\ref{eq:2PI}). Note that if one considers each of the contributions within brackets separately, they are all divergent. However, if we consider all terms simultaneously, we have
\beq
\left[\ln G^{-1}-\ln (G_0R)^{-1}+(G_0R)^{-1}\cdot G-1\right]\sim \frac{1}{2}(\Pi\,G_0R)^2\,,
\eeq
where we introduced $\Pi\equiv G^{-1}-(G_0R)^{-1}$. Because, in the UV, $\Pi$ receives only contribution from the tadpole integral, it follows that the trace contribution is convergent for appropriate choices of the regulator $R$.\\

In practice, it is convenient to consider the change of variables $G\rightarrow GR$. The gap equation then becomes
\beq\label{eq:2PIgap2}
\bar G^{-1}[\phi]=G_0^{-1}+2R\!\left.\frac{\delta\Gamma_{\rm int}}{\delta G}\right|_{\phi,\bar G[\phi] R}\,.
\eeq
This is the starting point that we have considered in order to write Eq.~(\ref{eq:2loopeq}) with the particular choice $R(Q)=\Theta(\Lambda-|Q|)$. Remember finally that the trace contribution in Eq.~(\ref{eq:2PI}) does not need regularization if all the terms are combined before taking the trace. One can however regularize the trace as 
\beq
{\rm Tr}\,\tilde R\left[\ln G^{-1}-\ln (G_0R)^{-1}+(G_0R)^{-1}\cdot G-1\right].
\eeq
This is useful for practical purposes because one can thus treat each term of the trace separately. Moreover, the gap equation (\ref{eq:2PIgap2}) becomes then
\beq
\tilde R\,\bar G^{-1}[\phi]=\tilde R\,G_0^{-1}+2R\!\left.\frac{\delta\Gamma_{\rm int}}{\delta G}\right|_{\phi,\bar G[\phi] R}\,.
\eeq
In particular, the choice $\tilde R=R$ shows that the presence of the regulator $R$ in front of the self-energy in Eq.~(\ref{eq:2PIgap2}) is not crucial for the matter of regularization.

\section{Subleading behavior of the bare self-energy}\label{sec:subleading}
In order to determine the next term in the asymptotic expansion of $\bar M^2(K)$ at large $\Lambda$, we subtract $\tilde M^2_\infty$, in the form given by Eq.~(\ref{eq:minf}), from Eq.~(\ref{eq:two}) with $L=0$. We obtain
\beq
\tilde M^2(\tilde K)-\tilde M^2_\infty & = & \tilde m_0^2+\frac{\lambda_0}{2}\,\tilde\phi^2+\frac{\lambda_0}{2}\int_{|\tilde Q|<1}\Big[\tilde G(\tilde Q)-\tilde G_\infty(\tilde Q)\Big]\nonumber\\
& - & \frac{\lambda_0^2}{2}\,\tilde\phi^2\mathop{\int_{|\tilde Q|<1}}_{|\tilde Q-\tilde K|<1}\tilde G_\infty(\tilde Q)\tilde G_\infty(\tilde Q-\tilde K)\,,
\eeq
where $\tilde G_\infty(\tilde Q)=1/(\tilde Q^2+\tilde M^2_\infty)$. Up to order $1/\Lambda^3$ contributions, one can write
\beq\label{eq:bebe}
\tilde M^2(\tilde K)-\tilde M^2_\infty & = & \tilde m_0^2+\frac{\lambda_0}{2}\,\tilde\phi^2-\frac{\lambda_0}{2}\int_{|\tilde Q|<1}\Big[\tilde M^2(\tilde Q)-\tilde M^2_\infty\Big]\tilde G_\infty^2(\tilde Q)\nonumber\\
& - & \frac{\lambda_0^2}{2}\,\tilde\phi^2\mathop{\int_{|\tilde Q|<1}}_{|\tilde Q-\tilde K|<1}\tilde G_\infty(\tilde Q)\tilde G_\infty(\tilde Q-\tilde K)+{\cal O}\left(\frac{1}{\Lambda^3}\right)\,.
\eeq
Note that the first integral in Eq.~(\ref{eq:bebe}) generates an infinite number of $1/\Lambda^2$ contributions. To resum them, we bring this term to the l.h.s. in the form (recall that $|\tilde K|<1$)
\beq\label{eq:sept}
& & \int_{|\tilde Q|<1}\left[\delta(\tilde K-\tilde Q)+\frac{\lambda_0}{2}\tilde G^2_\infty(\tilde Q)\right](\tilde M^2(\tilde Q)-\tilde M^2_\infty)\nonumber\\
& & \hspace{2.0cm}=\tilde m_0^2+\frac{\lambda_0}{2}\,\tilde\phi^2-\frac{\lambda_0^2}{2}\,\tilde\phi^2\mathop{\int_{|\tilde Q|<1}}_{|\tilde Q-\tilde K|<1}\tilde G_\infty(\tilde Q)\tilde G_\infty(\tilde Q-\tilde K)+{\cal O}\left(\frac{1}{\Lambda^3}\right)\,.
\eeq
The operator that appears in the l.h.s. is invertible since
\beq
\int_{|\tilde K|<1}\left[\delta(\tilde L-\tilde K)-\frac{\lambda_\infty}{2}\tilde G^2_\infty(\tilde K)\right]
\left[\delta(\tilde K-\tilde Q)+\frac{\lambda_0}{2}\tilde G^2_\infty(Q)\right]=\delta(\tilde L-\tilde Q)
\label{eq:inversion}
\eeq
holds if one chooses $1/\lambda_\infty=1/\lambda_0+(1/2)\int_{|\tilde Q|<1}\tilde G^2_\infty(\tilde Q)$. Applying the inverse operator on each side of Eq.~(\ref{eq:sept}), multiplying by $\Lambda^2$, renaming $\tilde L=K/\Lambda$ and using the fact that $K$ is kept fixed, we end up with Eq.~(\ref{eq:subleading}).

\section{Mean-field approximation for the mass-renormalized gap equation}\label{sec:mass_ren}
Let us here discuss the behavior as $\Lambda$ increases of the solution of the mean-field equations (\ref{eq:toy2}) and (\ref{eq:toy4}). 

\subsection{Equation for the zero-momentum self-energy}
The mean field approximation for the zero momentum self-energy is given by Eq.~(\ref{eq:g}). Note that $g_0(M^2)=-M^2+m^2+(\lambda_0/2)\phi^2$ and thus, for $\Lambda=0$, the equation $0=g_0(\bar M^2)$ admits one solution only: $\bar M^2=m^2+(\lambda_0/2)\phi^2$. In order to discuss the solutions of $0=g_\Lambda(\bar M^2)$ as one increases $\Lambda$ from $0$ to $\infty$, we study the profile of the function $g_\Lambda(M^2)$ as $\Lambda$ is varied. Note that its first and second derivatives with respect to $M^2$ are given by:
\beq
g'_\Lambda(M^2)=-1-\frac{\lambda_0}{32\pi^2}\left[\ln\frac{\Lambda^2+M^2}{M^2}-\frac{\Lambda^2}{\Lambda^2+M^2}\right]+\frac{\lambda_0^2\phi^2}{32\pi^2}\frac{\Lambda^4}{M^2(\Lambda^2+M^2)^2},\label{eq:gp}
\eeq
and
\beq
g''_\Lambda(M^2)=\frac{\lambda_0}{32\pi^2}\frac{\Lambda^4}{M^2(\Lambda^2+M^2)^2}\left[1-\frac{\lambda_0\phi^2}{M^2}\frac{\Lambda^2+3M^2}{\Lambda^2+M^2}\right].\label{eq:gpp}
\eeq
It is easily checked that $g''_\Lambda(M^2)$ changes sign only once. It starts infinitely negative as $M^2\rightarrow 0^+$, changes sign at some point and then remains positive, approaching $0$ as $M^2\rightarrow\infty$. It follows that $g'_\Lambda(M^2)$ decreases from $g'_\Lambda(0^+)=\infty$ to a certain value and then increases towards $g'_\Lambda(\infty)=-1$. This means that $g'_\Lambda(M^2)$ vanishes and changes sign only once for some $M^2=\bar M^2_{\rm e}(\Lambda)$. Correspondingly $g_\Lambda(M^2)$ increases from $g_\Lambda(0^+)=-\infty$ to $g_\Lambda(\bar M^2_{\rm e}(\Lambda))$ and then decreases to $-\infty$. The situation seems thus pretty similar to the one concerning the function $f_\Lambda(M^2)$ discussed in the text: the existence of solutions to the equation $0=g_\Lambda(\bar M^2)$ depends on the sign of $g_\Lambda(\bar M^2_{\rm e}(\Lambda))$. However the discussion of the sign of $g_\Lambda(\bar M^2_{\rm e}(\Lambda))$ is a little bit more involved than that of $f_\Lambda(\bar M^2_{\rm e}(\Lambda))$, if we want to do it for each value of $\Lambda$. Fortunately, we are only interested in what happens at higher values of $\Lambda$ and the discussion becomes again simple.\\

Consider first the equation
\beq
\frac{d}{d\Lambda}\bar M^2_{\rm e}(\Lambda)=-\frac{1}{g''_\Lambda}\frac{\partial g'_\Lambda}{\partial\Lambda}=-\frac{1}{16\pi^2 g''_\Lambda}\frac{\Lambda^3}{(\Lambda^2+\bar M^2_{\rm e}(\Lambda))^2}\left[\frac{2\lambda_0\phi^2}{\Lambda^2+\bar M^2_{\rm e}(\Lambda)}-1\right].
\eeq
For $\Lambda$ large enough, the r.h.s. becomes negative. It follows that $\bar M^2_{\rm e}(\Lambda)$ decreases with $\Lambda$ and thus that $\bar M^2_{\rm e}(\Lambda)$ converges as $\Lambda\rightarrow\infty$. From the definition $0=g'_\Lambda(\bar M^2_{\rm e}(\Lambda))$ and the explicit expression (\ref{eq:gp}), it follows that the only possible limit for $\bar M^2_{\rm e}(\Lambda)$ as $\Lambda\rightarrow\infty$ is $0$. Then, multiplying the equation $0=g'_\Lambda(\bar M^2_{\rm e}(\Lambda))$ by $\bar M^2_{\rm e}(\Lambda)$ and taking the limit $\Lambda\rightarrow\infty$, we obtain $\bar M^2_{\rm e}(\Lambda)\ln(\Lambda^2)\rightarrow \lambda_0\phi^2$, that is
\beq
\bar M^2_{\rm e}(\Lambda)\sim\frac{\lambda_0\phi^2}{\ln\Lambda^2}\,\,\,{\rm as}\,\,\,\Lambda\rightarrow\infty\,.
\eeq
We can now evaluate $g_\Lambda(\bar M^2_{\rm e}(\Lambda))$ from Eq.~(\ref{eq:g}) and study its behavior as $\Lambda\rightarrow\infty$. We obtain
\beq
g_\Lambda(\bar M^2_{\rm e}(\Lambda))\sim\frac{\lambda_0}{32\pi^2}(m^2-\lambda_0\phi^2)\ln\Lambda^2\,,
\eeq
if $m^2\neq \lambda_0\phi^2$, and 
\beq
g_\Lambda(\bar M^2_{\rm e}(\Lambda))\sim -\frac{\lambda_0^2\phi^2}{32\pi^2}\ln (\ln\Lambda^2)\,,
\eeq
if $m^2=\lambda_0\phi^2$. As in the case of $f_\Lambda(\bar M^2_{\rm e}(\Lambda))$, the value of $g_\Lambda(\bar M^2_{\rm e}(\Lambda))$ at large $\Lambda$ is driven by the presence of logarithmic divergences in the equation. But in contrast to $f_\Lambda(\bar M^2_{\rm e}(\Lambda))$, the sign of $g(\bar M^2_{\rm e}(\Lambda))$ at large $\Lambda$ now depends on the choice of parameters:\\

1. If $m^2\leq\lambda_0\phi^2$, $g_\Lambda(\bar M^2_{\rm e}(\Lambda))$ becomes negative above some $\Lambda_{\rm c}$ and the equation $0=g_\Lambda(\bar M^2)$ has no solution. This situation is similar to the one of equation $0=f_\Lambda(\bar M^2)$. The singular behavior $|\bar M^2-\bar M^2_{\rm c}|\propto (\Lambda_{\rm c}-\Lambda)^{1/2}$ in the vicinity of $\Lambda_{\rm c}$ is explained as in this later case.\\

2. If $m^2>\lambda_0\phi^2$, $g_\Lambda(\bar M^2_{\rm e}(\Lambda))$ is strictly positive for large values of $\Lambda$. This means that the equation $0=g_\Lambda(\bar M^2)$ has always two solutions for arbitrary large values of $\Lambda$. As we now show, the rightmost solution does not diverge as $\Lambda\rightarrow\infty$. Rather, it converges to the trivial limit $m^2-\lambda_0\phi^2$. To see this let us evaluate $g_\Lambda(M^2)$ for $M^2=(m^2-\lambda_0\phi^2)\pm\Delta m^2$ with $\Delta m^2>0$. Due to the presence of the logarithms
\beq
g_\Lambda(m^2-\lambda_0\phi^2\pm\Delta m^2)\sim\mp\frac{\lambda_0}{32\pi^2}\Delta m^2\ln \Lambda^2\,.
\eeq
It follows that there exists a value of $\Lambda$ above which $g_\Lambda(m^2-\lambda_0\phi^2-\Delta m^2)>0$ and $g_\Lambda(m^2-\lambda_0\phi^2+\Delta m^2)<0$. This means that, at least for $\Lambda$ large enough, the rightmost solution $\bar M$ of the equation $0=g_\Lambda(\bar M^2)$ \rt{has always two solutions, at least for $\Lambda$ large enough and that the rightmost solution} is such that $|\bar M^2-(m^2-\lambda_0\phi^2)|<\Delta m^2$. Because $\Delta m^2$ can be taken as small as desired, this shows that the rightmost solution admits a continuum limit: $\bar M^2\rightarrow m^2-\lambda_0\phi^2$. Using this information in Eq.~(\ref{eq:g}), we obtain
\beq
\bar M^2-(m^2-\lambda_0\phi^2)\sim \frac{1}{\ln\Lambda^2}\left((\lambda_0+48\pi^2)\phi^2+m^2\ln\frac{m^2-\lambda_0\phi^2}{m^2}\right),
\eeq
which shows that the limit is approached very slowly.

\subsection{Momentum dependent equations}
We can treat Eq.~(\ref{eq:toy4}) along similar lines. Notice first that it can be rewritten as $0=g_\Lambda(\bar M^2(K);K,\bar M^2)$ with
\beq
g_\Lambda(M^2;K,\bar M^2) & \equiv & -M^2+\bar M^2-\frac{\lambda_0^2}{2}\,\phi^2\left[\mathop{\int_{|Q|<\Lambda}}_{|Q+K|<\Lambda}\frac{1}{Q^2+\bar M^2}\frac{1}{(Q+K)^2+M^2}-\int_{|Q|<\Lambda}\frac{1}{(Q^2+\bar M^2)^2}\right].
\eeq
The first and second derivatives with respect to $M^2$ read
\beq
g'_\Lambda(M^2;K,\bar M^2) & = & -1+\frac{\lambda_0^2}{2}\,\phi^2\mathop{\int_{|Q|<\Lambda}}_{|Q-K|<\Lambda}\frac{1}{Q^2+\bar M^2}\frac{1}{((Q-K)^2+M^2)^2}\\
g''_\Lambda(M^2;K,\bar M^2) & = & -\lambda_0^2\,\phi^2\mathop{\int_{|Q|<\Lambda}}_{|Q-K|<\Lambda}\frac{1}{Q^2+\bar M^2}\frac{1}{((Q-K)^2+M^2)^3}\,.
\eeq
The second derivative being negative $g'_\Lambda(M^2;K,\bar M^2)$ decreases strictly with $M^2$ from $g'_\Lambda(0;K,\bar M^2)$ to $g'_\Lambda(\infty;K,\bar M^2)=-1$. If $|K|$ is small enough $g'_\Lambda(0;K,\bar M^2)$ is positive and $g_\Lambda(M^2;K,\bar M^2)$ increases first from $g_\Lambda(0;K,\bar M^2)$ up to a maximal value and then decreases to $g_\Lambda(\infty;K,\bar M^2)=-\infty$. For larger values of $|K|$, $g'(0;K,\bar M^2)$ could be negative and then $g_\Lambda(M^2;K,\bar M^2)$ would decrease all the way from $g(0;K,\bar M^2)$ to $g_\Lambda(\infty;K,\bar M^2)=-\infty$. Notice now that 
\beq
g(\bar M^2;K,\bar M^2)=\frac{\lambda_0^2}{2}\phi^2\Big[B(K)-B(0)\Big]\,,
\eeq 
where $B(K)$ is the bubble integral, defined in the next section, where the mass in the propagators is $\bar M^2$. We show in the next section that $B(K)$ is a monotonously increasing function of $|K|$. It follows that $g(\bar M^2;K,\bar M^2)$ is positive. Then, because for $m^2>\lambda_0\phi^2$ $\bar M^2$ is defined for arbitrary large values of $\Lambda$, it follows that Eq.~(\ref{eq:toy4}) has at least one solution for arbitrary large values of $\Lambda$. Equation (\ref{eq:toy4}) is compatible with this solution having a limit as $\Lambda\rightarrow\infty$. This limit $\bar M^2_\infty(K)$ obeys the equation (\ref{eq:limit}).

\section{About the evaluation of integrals}\label{sec:evaluation}

\subsection{The bubble integral with momentum dependent self-energy\label{ss:bubble}}
We begin with the main integral appearing throughout this work, the bubble integral defined as
\beq
B(K)=-\mathop{\int_{|Q|<\Lambda}}_{|Q-K|<\Lambda} \bar G(Q) \bar G(Q-K),
\label{Eq:bubble_def}
\eeq
where $\bar G(Q)=1/(Q^2+\bar M^2(Q)).$ Working with a 4D spherical coordinate system in which the angle between the vectors $Q$ and $K$ is $\theta,$ one can do the integrals over the remaining two angles  and introduce the variable $l^2=q^2+k^2-2k q\cos(\theta),$  where $q=|Q|$ and $k=|K|$ to write
\beq
B(k)=-\frac{1}{8\pi^3 k^2}\int_0^\Lambda d q\, f(q) \int_{|q-k|}^{{\rm min}(q+k,\Lambda)}d l\, g(l;q,k),
\label{Eq:bubble1}
\eeq
with $f(q)=q \bar G(q)$ and $g(l;q,k)=l \bar G(l) \sqrt{(l^2-(q-k)^2)((q+k)^2-l^2)}.$ We found that an efficient and robust way of computing numerically this integral is achieved upon using the change of variables\footnote{First the $l$-integral is converted into an integral on $[-1,1]$ and then on the entire real axis through the Tanh-Sinh transformation, which is commonly used in the so-called Tanh-Sinh quadrature, see for instance \cite{tanh-sinh1,tanh-sinh2}.} $l=u(y;q)\equiv \frac{1}{2}(b(q)+a(q))+\frac{1}{2}(b(q)-a(q))\tanh(\frac{\pi}{2} \sinh(y)),$ where $b(q)={\rm min}(q+k,\Lambda),$ $a(q)=|q-k|$ followed by $q=v(x)\equiv \frac{\Lambda}{2}\left[1+\tanh\left(\frac{\pi}{2}\sinh(x)\right) \right].$ After these transformations the bubble integral reads
\beq
B(k)=-\frac{\Lambda}{32\pi k^2}\int_{-\infty}^\infty d x \, J(x) (b(v(x))-a(v(x))) f(v(x)) \int_{-\infty}^\infty d y \, J(y) g(u(y;v(x));v(x),k),
\eeq
where $J(z)=\frac{\cosh(z)}{\cosh^2(\frac{\pi}{2}\sinh(z))}.$ The great advantage is that after these transformations both integrands decays with a double exponential rate towards both ends of the real axis, and in all practical situations it is enough to integrate over a relatively narrow compact interval, {\it e.g.} $[-4,4],$ within which the integrands have a maximum.  In some cases the adaptive numerical integration routines of the GNU Scientific Library (GSL)~\cite{gsl} crashed due to the small values of the functions. In order to avoid this, these integrals were approximated by the integrals of the corresponding Chebyshev approximated functions in which the Chebyshev coefficients where calculated using a fast cosine transform routine of the Fastest Fourier Transform in the West (FFTW) library \cite{FFTW3}. We varied the number of Chebyshev polynomials between 150 and 350 and checked that the bubble integral is calculated with a relative error smaller than $10^{-8}$.\\

\subsection{The perturbative bubble integral and its monotonous behavior\label{ss:pert_bubble}}

\begin{figure}[t]                                                               
\begin{center}                                                                  
\resizebox{0.4\textwidth}{!}{\begin{tikzpicture}[set style={{help lines}+=[very thin]},>=stealth]
\colorlet{examplefill}{black!10!white}

\draw[very thin, dashed] (0,0) circle (0.7cm);
\draw[very thin, dashed] (0,0) circle (2.5cm);

\draw[color=white,thick] (0,0) circle (3cm);
\draw[color=white,thick] (1.7,0) circle (3cm);
\fill[fill=examplefill,even odd rule, opacity=1.0, path fading=fade out] (0,0) circle (3cm) (1.7,0) circle (3cm);

\draw [color=red,thick] (1.7,2.47184141886166) arc (124.518107841061:235.481892158939:3cm);
\draw [color=red,thick] (1.7,2.47184141886166) arc (55.4818921589387:-55.4818921589387:3cm);

\draw [color=black,thick] (0.85,2.877064476163) arc (106.45924962826:253.54075037174:3.0cm);
\draw [color=black,thick] (0.85,2.877064476163) arc(73.5407503717398:-73.5407503717398:3.0cm);
\draw [color=red,thick] (1.7,2.47184141886166) arc (55.4818921589387:-55.4818921589387:3cm);

\draw [color=red,thick](1.7,-2.47184141886166) -- (1.7,2.47184141886166);

\fill (0,0) circle (1.5pt) node[anchor=east]{$0$};
\fill (1.7,0) circle (1.5pt) node[anchor=west]{$K$};
\draw [->](0,0) -- (1.7,0); 
\draw (2.1,-1.4) node[color=red,rectangle, draw]{$D$};
\draw (1.25,-1.4) node[color=red,rectangle, draw]{$\bar D$};
\draw (0.2,-1.7) node[rectangle, draw]{$C$};

\node[right] at (2,1.5) {$Q$};
\node[left] at (1.3,1.5) {$\bar Q$};
\draw[very thin, dashed] (1.4,1.5) -- (2,1.5);

\draw[->] (0,0) -- (2,1.5);
\draw[->] (0,0) -- (1.4,1.5);

\draw[->] (1.7,0) -- (2,1.5);
\draw (1.7,0) -- (1.4,1.5);

\draw [->,double distance=0.4pt] (1,0) arc (0:37.7:1); 
\draw [->,] (0.35,0) arc (0:50.0:0.35); 
\node[right] at (0.95,0.35) {{\small $\theta$}};
\node[right] at (0.3,0.2) {{\small $\bar\theta$}};

\end{tikzpicture}}                               
\caption{Region of integration for the bubble integral. The dashed circle shows a case in which $\theta\in[0,\pi],$ while the dashed arc corresponds to a case when $0<\theta<\arccos\,\alpha.$ The remaining part of the plot shows the construction used to prove that the bubble integral is a monotone increasing function of the momentum. 
\label{Fig:spheres}}                                                            
\end{center}                                                                    
\end{figure}
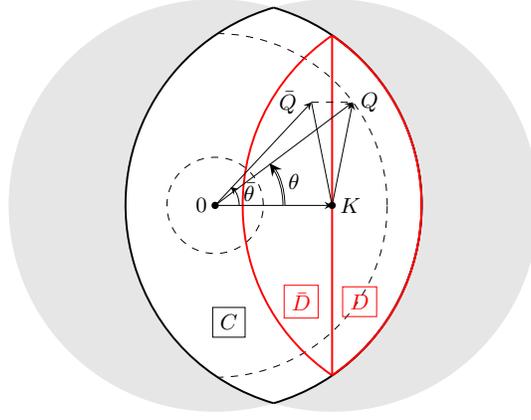

To compute the bubble integral defined with a propagator containing a momentum independent self-energy, that is $\bar G(Q)=1/(Q^2+\bar M^2),$ we use the following geometrical picture (see also Fig.~\ref{Fig:spheres}).  With a 4D spherical coordinate system we have to integrate over the common region of two spheres\footnote{This region is always nonvanishing if $|Q|, |K|, |Q-K|<\Lambda.$} of radius $\Lambda$ with distance $k\equiv|K|$ between their centers. Then, the integral over two angles gives $4\pi,$ while, depending on the value of $q\equiv|Q|,$ the remaining angle $\theta$ can go from $0$ up to $\pi$ or to $\arccos\,\alpha<\pi$, with $\alpha\equiv (k^2+q^2-\Lambda^2)/(2k q)$ determined by the intersection ``point'' of the two spheres.  Then, the integral of (\ref{Eq:bubble_def}) reads:
\beq
B_{\rm pt}(k) & = & -\frac{1}{4\pi^3}\left[\int_0^{\Lambda-k}d q\, \frac{q^3}{q^2+\bar M^2} T(q,k,\bar M,-1) + \int_{\Lambda-k}^\Lambda d q\, \frac{q^3}{q^2+\bar M^2} T(q,k,\bar M,\alpha)\right],
\eeq
where 
\beq
T(q,k,\bar M,\alpha) & \equiv & \int_0^{\arccos\,\alpha} d\theta\,\frac{\sin^2\theta}{q^2+k^2-2q\,k\cos\theta + \bar M^2}.
\label{Eq:T_def}
\eeq

\noindent
The $\theta$-integral above can be performed analytically with the change of variable $t=\tan(\theta/2):$
\beq
T(q,k,\bar M,\alpha) & = & \frac{8}{(q+k)^2+\bar M^2}\int_0^{t_{\rm max}} d t \frac{t^2}{(1+t^2)^2}\, \frac{1}{t^2+a^2} \nonumber\\
& = & \frac{1}{4q^2k^2}\left[ 2q\,k\sqrt{1-\alpha^2}+\big(q^2+k^2+\bar M^2\big)\arccos\,\alpha -2\varepsilon_+\varepsilon_-\arctan\,\left(
\frac{\varepsilon_+}{\varepsilon_-}\sqrt{\frac{1-\alpha}{1+\alpha}}
\right)
\right],\ \ 
\label{Eq:T_general}
\eeq
where we introduced $\displaystyle t_{\rm max}=\frac{1-\alpha}{1+\alpha},$  $\displaystyle a^2=\frac{(q+k)^2+\bar M^2}{(q-k)^2+\bar M^2},$ and $\varepsilon_{\pm}=\sqrt{(q\pm k)^2+\bar M^2}$. Note that, as special cases one has  
\beq
T(q,k,\bar M,-1) & = &\frac{\pi}{4q^2k^2} \Big[q^2+k^2+\bar M^2-\varepsilon_+\varepsilon_-\Big] = \frac{2\pi}{(\varepsilon_++\varepsilon_-)^2}\,,\nonumber\\
T(q,k,0,-1) & = & \frac{\pi}{2}{\rm Max}(q,k)^{-2}\,.
\label{Eq:T_spec}
\eeq

Using a purely geometric argument one can prove a useful property of $B_{\rm pt}(K),$ namely that it increases with $|K|.$ To do so, one considers the derivative of the bubble integral with respect to $ K|,$
\beq
\frac{d B_{\rm pt}(K)}{d|K|} & = &2 I_1(K)+2 I_2(K),
\nonumber\\
I_1(K)&=&\mathop{\int_{|Q|<\Lambda}}_{|Q-K|<\Lambda} \frac{1}{Q^2+\bar M^2}\frac{|K|-|Q|\cos\theta}{((Q-K)^2+\bar M^2)^2},
\nonumber\\
I_2(K)&=&\int_{|Q|<\Lambda}\frac{|K|-|Q|\cos\theta}{Q^2+\bar M^2} \frac{\delta(\Lambda^2-(Q-K)^2)}{(Q-K)^2+\bar M^2}\,,
\eeq
and prove that it increases.  Let us treat first the contribution $I_1(K)$. The trick is to decompose the integration domain $\{|Q|<\Lambda\}\cap\{|Q-K|<\Lambda\}$ in three disjoint domains $C$, $D$ and $\bar D$, see Fig.~\ref{Fig:spheres}. The region $D$ is $\{|Q|<\Lambda\}\cap\{|Q-K|<\Lambda\}\cap\{|Q|\cos\theta > |K|\}$. The region $\bar D$ is the mirror symmetric of $D$ with respect to the axis $|Q|\cos\theta=|K|$. The region $C$ is $\{|Q|<\Lambda\}\cap\{|Q-K|<\Lambda\}\backslash (D\cup\bar D)$. One has $I_1=I_C+I_D+I_{\bar D}$. In region $C$ (and also in region  $\bar D$), one has $|K|-|Q|\cos\theta>0$, from which it follows that $I_C>0$. In order to treat $I_D$ and $I_{\bar D}$, for each point $Q$ in region $D$, we introduce its mirror symmetrized  $\bar Q=\bar Q(Q)$. We denote by $\bar\theta=\bar\theta(Q)$ the  corresponding angle. It is easily checked on Fig.~\ref{Fig:spheres} that for each $Q\in D$, $|K|-|\bar Q|\cos\bar\theta=-(|K|-|Q|\cos\theta)>0$,  $|Q-K|=|\bar Q-K|$ and $|\bar Q|<|Q|$. We can now write
\beq
I_D+I_{\bar D}=\int_{Q\in D}\frac{|K|-|Q|\cos\theta}{((Q-K)^2+\bar M^2)^2} \left[\frac{1}{Q^2+\bar M^2}-\frac{1}{\bar Q^2(Q)+\bar M^2}\right]>0\,.
\eeq
Next, let us consider the contribution $I_2,$ which corresponds to the variation of $B_{\rm pt}(K)$ due to a change of the integration domain determined by the intersection of the two spheres (see Fig.~\ref{Fig:spheres}). One can see geometrically that when $|K|$ increases, the two spheres separate apart and the volume of the integration domain decreases. Therefore, the positive value of the integral decreases leading to the increase of $B_{\rm pt}(K).$ This argument prove the positivity of $I_2,$ which can be checked by a simple calculation.  The support of the delta function is contained within the integration domain if $-1<\frac{q^2+k^2-\Lambda^2}{2q k}<1$ that is $|q-k|<\Lambda$ and $q+k>\Lambda.$ For $q,k<\Lambda$ the first constraint is clearly satisfied, so that one has
\beq
I_2(k)& = & \frac{1}{32\pi^2 k^2}\frac{1}{\Lambda^2+\bar M^2} \int_{\Lambda-k}^{\Lambda}d q\,q^2\,\frac{\Lambda^2-q^2+k^2}{q^2+\bar M^2}.
\eeq
Now, since $\Lambda^2+k^2-q^2$ is positive at both ends of the $q$-interval and as a function of $q$ it is strictly decreasing, it follows that it is positive over the whole interval. This means that $I_2$ is positive, which completes the proof. 

We found numerically that this property of the perturbative bubble integral is inherited by the solution $\bar M(K)$ of the self-consistent gap equation containing $B(K)$, which eventually becomes a monotone increasing function of $K.$

\subsection{The double integrals of Eq.~(\ref{eq:complete}) \label{ss:double_int}}
Next, we discuss the two double integrals appearing in Eq.~(\ref{eq:complete}). After shifting the variable of integration $Q\to Q-R,$ one switches the order of the integrals and uses the method given in (\ref{ss:pert_bubble}). For the first double integral one obtains 
\beq  
C&=&\int_{|R|<\Lambda}\bar G(R)\mathop{\int_{|Q|<\Lambda}}_{|Q-R|<\Lambda} \bar G(Q) G_0^2(Q-R)=-\frac{1}{32\pi^5}\int_0^\Lambda d r\, r^3 \bar G(r)\nonumber\\
&&\qquad \times\left[ \int_0^{\Lambda-r} d q\, q^3 \bar G(q)\frac{d }{d m^2}T(q,r,m,-1)+\int_{\Lambda-r}^\Lambda d q\, q^3 \bar G(q)\frac{d }{d m^2} T(q,r,m,\alpha) \right],
\label{Eq:C_def}
\eeq
where $G_0(Q)=1/(Q^2+m^2)$ and $\alpha=(r^2+q^2-\Lambda^2)/(2r q)$ with $r\equiv|R|$ and $q\equiv|Q|.$ Using (\ref{Eq:T_general}) and (\ref{Eq:T_spec}) the derivatives can be readily done. Then, for
the $Q$-integrals we do a Tanh-Sinh transformation as in (\ref{ss:bubble}) and evaluate the integrals numerically. The second double integral involves only the propagator $G_0$ and is defined as
\beq
C_0&=&\int_{|R|<\Lambda}G_0(R)\mathop{\int_{|Q|<\Lambda}}_{|Q-R|<\Lambda} G_0(Q) G_0^2(Q-R).
\label{Eq:C0_def}
\eeq
It can be evaluated exactly as $C,$ the $Q$-integral with $\alpha=-1$ can be even done analytically.\\

We shall denote by $C_{\rm pt}$ the double integral similar to $C,$ but defined with momentum independent self-energy, that is with $\bar G(Q)=1/(Q^2+\bar M^2).$ Its divergence can be obtained by studying
the divergence of its derivative with respect to $\Lambda.$ Changing the order of the integration in the original form of the integral given in (\ref{eq:complete}) one has
\beq
\frac{d C_{\rm pt}}{d \Lambda} &=& \int_{|R|<\Lambda} \bar G(R) \int_{|Q-R|<\Lambda} G_0^2(Q) \bar G(Q-R)\delta(\Lambda-|Q|)\nonumber \\
&&+2\int_{|R|<\Lambda} \delta(\Lambda-|R|) \bar G(R) \mathop{\int_{|Q|<\Lambda}}_{|Q-R|<\Lambda} G_0^2(Q) \bar G(Q-R)\, .\ \ 
\label{Eq:dCL_def}
\eeq
Using in both terms the method leading to (\ref{ss:pert_bubble}) one obtains
\beq
\frac{d C_{\rm pt}}{d \Lambda} =\frac{1}{32\pi^5}\left[\frac{\Lambda^3}{(\Lambda^2+m^2)^2}\int_0^\Lambda d r\, r^3 \bar G(r) T(r,\Lambda,\bar M,\alpha)+\frac{2\,\Lambda^3}{\Lambda^2+\bar M^2}
\int_0^\Lambda d q\, q^3 G_0^2(q) T(q,\Lambda,\bar M,\beta)\right],\qquad
\label{Eq:dCL1}
\eeq
where $\alpha=r/(2\Lambda)$ and $\beta=q/(2\Lambda).$
Rescaling everything by $\Lambda,$ one introduces $\tilde m^2=m^2/\Lambda^2$
and $\tilde M^2=\bar M^2/\Lambda^2$ to write
\beq
\frac{d C_{\rm pt}}{d \Lambda} = \frac{1}{128\pi^5 \Lambda}\left[\frac{1}{(1+\tilde m^2)^2}\int_0^1 d u\frac{u}{u^2+\tilde M^2}f(u;\tilde M^2)+\frac{2}{(1+\tilde M^2)}\int_0^1 d u\frac{u}{(u^2+\tilde m^2)^2}f(u;\tilde M^2)\right],
\label{Eq:dCL2}
\eeq
with
\beq
f(u;\tilde M^2)&=&u\sqrt{4-u^2}+(u^2+\tilde M^2+1)\arccos\left(\frac{u}{2}\right)\nonumber\\
&&-2\sqrt{(u^2-1)^2+2\tilde M^2(u^2+1)+\tilde M^4} \arctan\left(\sqrt{\frac{2-u}{2+u}}\sqrt{\frac{(u+1)^2+\tilde M^2}{(u-1)^2+\tilde M^2}}\,\right).
\eeq
We are interested in the behavior of the integrals in (\ref{Eq:dCL2}) at small $\tilde M^2$ and $\tilde m^2$, but because for small $u$ one has $f(u;\tilde M^2) \simeq \pi u^2/(1+\tilde M^2),$ one can set 
$\tilde m^2=\tilde M^2=0$ in the second integral of (\ref{Eq:dCL2}) only after subtracting from the integrand its leading contribution at  small $u,$ otherwise the integral would diverge in the infrared. Adding and subtracting this leading contribution and setting $\tilde m^2=\tilde M^2=0$ whenever it is possible, one finds
\beq
\frac{d C_{\rm pt}}{d \Lambda} = \frac{1}{128\pi^5\Lambda}\left\{\int_0^1 \frac{d u}{u} \left[f(u;0)\left(1+\frac{2}{u^2}\right)-2\pi\right]+2\pi\int_0^1d u \frac{u^3}{(u^2+\tilde m^2)^2}\right\}
+{\cal O}\left(\frac{\ln\Lambda}{\Lambda^2}\right).
\eeq
Using the symbolic manipulation program {\it Mathematica} the first integral is evaluated to $\pi$~(!), while the second integral can be done analytically, and for small $\tilde m$ its leading contribution is $-\big(1+\ln\tilde m^2\big)/2.$ Therefore, the divergence of $d C_{\rm pt}/d \Lambda$ is given by $(1/64\pi^4\Lambda)\ln(\Lambda/m)$,  which in turn means that the divergence of $C_{\rm pt}$ is $(1/128\pi^4)\ln^2(\Lambda/m)$, independently on $\bar M$.

\section{Numerical solution of the gap and flow equations \label{sec:flow}}

The numerical algorithm used to solve the gap equation is illustrated on the completely renormalized version of this equation, that is Eq.~(\ref{eq:complete}). It is useful to rewrite this equation  in the following form 
\beq
\bar M^2(k)=m^2+\frac{\lambda}{2}\phi^2 \left(1+\lambda B_0-\frac{\lambda^2}{2} C_0\right)+\frac{\lambda}{2}\left(D+\frac{\lambda^2}{2}\phi^2 C\right)-\frac{\lambda^2}{2}\phi^2 B(k),
\label{Eq:complete_num}
\eeq
where $k=|K|$ and
\beq
B_0&=&\int_{|Q|<\Lambda}G^2_0(Q)=\frac{1}{16\pi^2} \left[\ln\left(\frac{\Lambda^2+m^2}{m^2}\right)-\frac{\Lambda^2}{\Lambda^2+m^2}\right],\nonumber\\
D&=&\int_{|Q|<\Lambda}(\bar M^2(Q)-m^2)^2G^2_0(Q)\bar G(Q)=\frac{1}{8\pi^2}\int_0^\Lambda d q\, (\bar M^2(q)-m^2)^2 q^3 G_0^2(q) \bar G(q).
\eeq
The integrals $B(k),$ $C_0,$ and $C$ are defined in (\ref{Eq:bubble_def}), (\ref{Eq:C_def}), and (\ref{Eq:C0_def}). The function $\bar M^2(k)$ is stored on a non-equidistant grid, that is at discrete values of the momentum given by $\displaystyle k(j)=k_{\rm m}+(\Lambda-k_{\rm m}) \big(j/(N-1)\big)^\frac{5}{2},$ where $k_{\rm m}$ is the modulus of the smallest momentum stored, $N$ is the number of discretization points and $j=0,\dots,N-1.$ Note that the grid was chosen to be finer for small momenta. 

To solve the equation iteratively one starts with a constant value for $\bar M^2(k)$ determined by the terms on the right hand side of Eq.~(\ref{Eq:complete_num}) which contains the integrals $B_0$ and
$C_0$ (they are independent on $\bar M$), then at every new iteration the integrals $D,$ $C,$ and $B(K)$ are calculated using the stored values of $\bar M^2(K),$ and at the end the values of $\bar M^2(K)$  are upgraded using the right hand side of Eq.~(\ref{Eq:complete_num}).  One uses the one-dimensional Akima spline interpolation method to calculate the value of $\bar M^2(q)$ at arbitrary momenta, required by the integration routine used to evaluate the integrals. The numerical evaluation of this integrals was described in Appendix~\ref{sec:evaluation}. At a given value of the cut-off the number of discretization points used was $[100 \Lambda]$ or $[400 \Lambda/3].$\\

We present below the method used to solve the flow equations (\ref{eq:flow}) and (\ref{Eq:flow_complete}). Using the notations of subsection~\ref{ss:bubble} and working for the sake of simplicity with unscaled quantities, the flow equation (\ref{Eq:flow_complete}) can be rewritten in the following form
\beq
\int_0^\infty d q\big[\delta(k-q)-H(k,q)\big]\theta(\Lambda-q)\partial_\Lambda \bar M^2(q)=Z(k),
\label{Eq:flow_rewritten}
\eeq 
where 
\beq
H(k,q)&=&\frac{\lambda_0^2\phi^2}{8\pi^3 k^2\Lambda} h(k,q)q\bar G^2(q)-\frac{\lambda_0}{16\pi^2}q^3 \bar G^2(q),\nonumber\\
Z(k)&=&-\frac{\lambda_0^2\phi^2}{4\pi^3 k^2\Lambda} z(k)+\frac{\lambda_0 A_\Lambda}{8\pi^2\Lambda}.
\label{Eq:H_and_Z}
\eeq
Here $h(k,q)$ denotes the $l$-integral in (\ref{Eq:bubble1}) and we have also introduced
\beq
z(k)=\int_0^\Lambda d q\, q \bar G^2(q) \bar M^2(q) h(k,q),\qquad A_\Lambda=\int_0^\Lambda d q q^3 \left[\frac{q^2+2\bar M^2(q)}{(q^2+\bar M^2(q))^2}-\frac{q^2+2 m^2}{(q^2+m^2)^2}\right].
\label{Eq:z_and_A}
\eeq
The equation (\ref{eq:flow}) can be written in the same form as (\ref{Eq:flow_rewritten}), only that in both $H(k,q)$ and $Z(k)$ the second term is absent. In order to solve Eq.~(\ref{Eq:flow_rewritten})
numerically we store $M^2(q)$ on a non-equidistant grid and uses the one-dimensional Akima spline interpolation method to calculate $M^2(q)$ at arbitrary momentum value required by the integration  routine used to evaluate the integrals in (\ref{Eq:z_and_A}). These are the only integrals remaining since upon discretization Eq.~(\ref{Eq:flow_rewritten}) becomes
\beq
\sum_{j=0}^{N-1}\big[\delta_{ij}-\Delta_i H_{ij}\big]\frac{d \bar M^2_j}{d \Lambda}=Z_i,\label{Eq:flow_discrete}
\eeq
where $\Delta_i$ denotes the distance between two consecutive momenta on the non-equidistant grid. It is evident from (\ref{Eq:flow_discrete}) that one has to solve a system of $N$ ordinary differential equations (ODE) of the form  $d \bar M^2_j/d \Lambda=x_j(\bar M_0^2,\dots, \bar M_{N-1}^2, \Lambda),$ where in order to know $x_j$ the linear matrix equation $\sum_{j=0}^{N-1} \big[\delta_{ij}-\Delta_i H_{ij}\big] x_j=Z_i$ has to be solved at every step of the ODE solving algorithm. To solve the ODEs we used the implicit 4th order Runge-Kutta method, the matrix equation was solved using LU decomposition and the number of discretization points was $N=400$ or $N=500$ when solving Eq.~(\ref{Eq:flow_complete}) and $N=175$ when solving Eq.~(\ref{eq:flow}).

\end{document}